\documentclass[%
reprint,
superscriptaddress,
 amsmath,amssymb,
 aps,
prb,
]{revtex4-2}
\usepackage{graphicx}% Include figure files
\usepackage{siunitx}
\usepackage{dcolumn}% Align table columns on decimal point
\usepackage{bm}% bold math
\usepackage{hyperref}% add hypertext capabilities
\usepackage{xcolor}

\begin{document}

\title{Optical spectroscopy of single- and two-ion transitions in an antiferromagnetic stoichiometric rare-earth crystal}

\author{Masaya Hiraishi}

\author{Gabrielle A. Hunter-Smith}

\author{Gavin G. G. King}

\affiliation{Department of Physics, University of Otago, Dunedin New Zealand}
\affiliation{Dodd-Walls Centre for Photonic and Quantum Technologies, Dunedin, New Zealand}

\author{Alexandra A. Turrini}
\affiliation{Institute of Physics, Ecole Polytechnique F\'ed\'erale de Lausanne (EPFL), CH-1015 Lausanne, Switzerland}
\affiliation{Laboratory for Neutron Scattering and Imaging, Paul Scherrer Institut, 5232 Villigen PSI, Switzerland}
\author{J.-R. Soh}
\address{Quantum Innovation Centre (Q.InC), Agency for Science, Technology and Research (A*STAR), 2 Fusionopolis Way, Innovis \#08-03, Singapore 138634, Singapore}
\address{Institute of Materials Research and Engineering (IMRE), Agency for Science Technology and Research (A*STAR), 2 Fusionopolis Way, Innovis \#08-03, Singapore 138634, Republic of Singapore}
\address{Centre for Quantum Technologies, National University of Singapore, 3 Science Drive 2, Singapore 117543, Singapore}
\author{Henrik M. R\o{}nnow}
\affiliation{Institute of Physics, Ecole Polytechnique F\'ed\'erale de Lausanne (EPFL), CH-1015 Lausanne, Switzerland}
\author{Luke S. Trainor}
\author{Jevon J. Longdell}
\affiliation{Department of Physics, University of Otago, Dunedin New Zealand}
\affiliation{Dodd-Walls Centre for Photonic and Quantum Technologies, Dunedin, New Zealand}

\date{\today}

\begin{abstract}
We characterise optical transitions of neodymium ions (Nd\textsuperscript{3+}) in antiferromagnetic neodymium gallate (NdGaO\textsubscript{3}) with applied fields up to \qty{3}{\tesla}.
The magnetic phase of this material has not previously been studied with the field along its magnetisation axis.
The measured optical spectra indicate three magnetic phases\textemdash antiferromagnetic, intermediate, and paramagnetic\textemdash where the intermediate phase likely forms a different magnetic structure from typical spin-flop phases.
The observed absorptions were classified into two distinct families of optical transitions: single-Nd and two-Nd absorptions.
We demonstrate that the optical transitions in the antiferromagnetic and paramagnetic phases can be modelled using a standard single-ion crystal-field Hamiltonian that interacts with a mean magnetisation from the rest of the lattice, and we expand that model to encompass pairs of ions, explaining the origins of the two-Nd transitions.
This study offers a deeper understanding of the optical transitions in rare-earth antiferromagnetic crystals, which have been recently attracting significant interest for microwave-to-optical quantum transduction, despite being relatively unexplored to date.
\end{abstract}

\maketitle

\section{Introduction}
Rare-earth crystals offer outstanding spectroscopic properties in their $4f$-to-$4f$ energy-state transitions. Among them are narrow optical linewidths and long coherence times, which have been reported on various rare-earth ions in various crystals \cite{thiel_rare-earth-doped_2011, macfarlane_optical_1998, Fraval_2004, Fraval_2005, Bottger_2009, Lovric_2011, Zhong_2015, Rakonjac_2020}.
Those $4f$ transitions allow efficient storage, manipulation, and retrieval of quantum information with optical photons. Besides, spin and optical transitions enable microwave-to-optical frequency transduction, which is necessary for long-distance communication between superconducting qubits. To date, a wide variety of quantum applications have been theoretically proposed and experimentally demonstrated, such as quantum memories \cite{thiel_rare-earth-doped_2011, Guo_2023, azuma_quantum_2023} and transducers \cite{Williamson_2014, Fernandez_2015, Everts_2019, Fernandez-Gonzalvo_2019, Barnett_2020, Bartholomew_2020, King_2021, Xie_2021, Lim_2024}.

The vast majority of investigations into the coherent properties of rare-earth crystals have been for rare-earth dopants in non-magnetic crystals, typically with concentrations in the hundreds of parts per million or less. The low concentration reduces the decoherence caused by the interactions between rare-earth ions. The low concentration also results in less crystal distortion and narrower inhomogeneous linewidths.
Indeed, an optical coherence time of 4.4\,ms was achieved in Er:Y\textsubscript{2}SiO\textsubscript{5} \cite{Bottger_2006_optical}, which is, so far, the longest optical coherence time in any solid. A narrow linewidth of 10\,MHz was observed in Nd:Y\textsuperscript{7}LiF\textsubscript{4} \cite{macfarlane_optical_1998}.

Increasing the concentration of the dopants generally makes the linewidths broader. But if the `dopant’ concentration is increased in the extreme, to the point where the dopant is now part of the host crystal, many of the attractive properties seen at low concentrations reappear.
The rare-earth ions no longer distort the host crystal by their presence because they are part of the host crystal. Indeed, similar spectroscopic properties to the rare-earth-doped non-magnetic crystals were found in stoichiometric rare-earth non-magnetic crystals about a decade ago. An optical inhomogeneous linewidth of 25\,MHz was observed in Eu\textsuperscript{35}Cl\textsubscript{3}$\cdot$6H\textsubscript{2}O \cite{Ahlefeldt_2016}.
It was shown that coherence times could be prolonged to approximately \qty{740}{\us} in EuCl\textsubscript{3}$\cdot$6D\textsubscript{2}O \cite{ahlefeldt_optical_2013}. Those groundbreaking works have demonstrated that, as well as rare-earth-doped non-magnetic crystals, stoichiometric rare-earth non-magnetic crystals are a potential material for quantum applications.

Surprisingly, rare-earth ions in magnetic crystals have recently emerged as another area of interest, alongside non-magnetic crystals. Rare-earth non-magnetic crystals have been traditionally explored because thermal fluctuation of electronic spins is a dominant decoherence source \cite{Bottger_2006_optical}. Nevertheless, such thermal fluctuation can be suppressed at temperatures low enough that those spins are magnetically ordered. Indeed, optical transmission spectroscopy of an antiferromagnetic \textsuperscript{7}LiErF\textsubscript{4} crystal showed transitions with optical inhomogeneous linewidths of 18\,MHz \cite{Berrington_2022}, which is as narrow as \textsuperscript{170}Er:Y\textsuperscript{7}LiF\textsubscript{4} (16\,MHz in Ref.\,\cite{thiel_rare-earth-doped_2011}) and \textsuperscript{166}Er:Y\textsuperscript{7}LiF\textsubscript{4} (16\,MHz in Ref.\,\cite{kukharchyk_optical_2018}). Besides, a collective excitation of spins (a magnon) in magnetically ordered crystals can effectively couple to a microwave cavity mode, which holds promise for effective microwave-to-optical quantum transduction \cite{Everts_2019}. Recently a long optical coherence time was achieved in Er:GdVO\textsubscript{4} (\qty{235.7}{\us}) \cite{hiraishi_long_2025}, which is comparable with \textsuperscript{167}Er:YVO\textsubscript{4} (\qty{336}{\us}) \cite{li_optical_2020}.

Compared to non-magnetic rare-earth crystals, the properties of those magnetic crystals are relatively poorly described. One model is an effective Hamiltonian for \textsuperscript{7}LiErF\textsubscript{4} \cite{Berrington_2022}. There, the $4f$ energy-state transition frequencies depend on the spin configuration of a magnetically ordered state (e.g. an antiferromagnetic state).
The linewidths of the observed transitions are narrower than the strength of the magnetic interactions. This means that these spectroscopic studies also provide new insights into the magnetic ordering in these materials.

In this paper, we investigate optical transitions of an antiferromagnetic NdGaO\textsubscript{3} crystal under applied magnetic fields. During a series of measurements, the temperature was kept about 40\,mK, which is much lower than the ordering temperature of 0.97\,K. Optical transmission was measured to characterise optical transitions under applied magnetic fields. Two sets of measurements were made: one with the magnetic field along the magnetisation axis, and one with the field perpendicular.

\section{Methodology}
\subsection{\label{subsec: material} Material information: NdGaO\texorpdfstring{\textsubscript{3}}{₃}}
\begin{figure}[tb]
\begin{center}
\includegraphics[width=\linewidth]{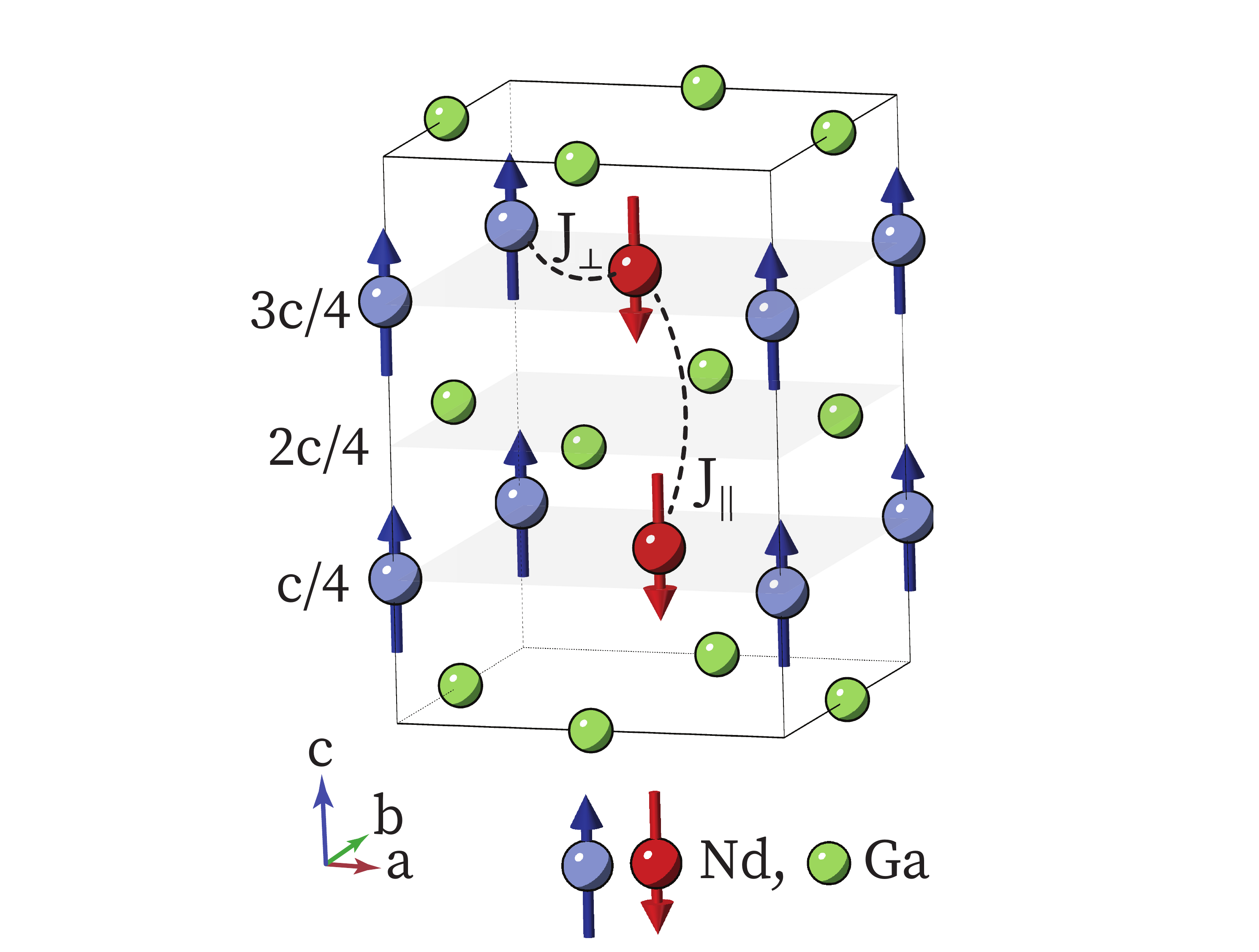}
\caption{\label{fig: Magnetic Structure} Magnetic structure of NdGaO$_3$ below the N\'eel temperature with zero external magnetic field, using the conventions in Refs.\,\cite{Marti1995, Luis1998}.}
\end{center}
\end{figure}

The crystalline structure of NdGaO$_3$ is orthorhombic, belonging to space group $D^{16}_{2h}$ ($Pbnm$) \cite{Marti1995}. There are four Nd$^{3+}$ ions within the unit cell, all of which are characterised by the $4c$ Wickoff positions with a $C_s$ ($m$) site symmetry. The lattice constants along $(a, b, c)$ axes are $(5.4223, 5.4994, 7.6989)$ \r{A} \cite{Marti1995}. NdGaO\textsubscript{3} becomes antiferromagnetically ordered below the N\'eel temperature of 0.97\,K with a $c_z$ configuration \cite{Marti1995, Luis1998}, as shown in Fig.\,\ref{fig: Magnetic Structure}.
Below this temperature, adjacent Nd spins on the same $ab$ plane (``in-plane'' spins) are anti-parallel, whereas out-of-plane spins are parallel. Luis \textit{et al.}, modelled the magnetic structure based on an assumption that there were two exchange interactions $J_{\parallel}$ between `out-of-plane' spins oriented parallel, and $J_{\perp}$ between `in-plane' spins oriented anti-parallel \cite{Luis1998}. There, the effective spin-1/2 Hamiltonian, $\hat{H}_{\mathrm{spin}\text{-}\frac{1}{2}}$, was modelled as
\begin{align}
\label{eq: Luis Hamiltonian}
        \hat{H}_{\mathrm{spin} \text{-}\frac{1}{2}} =& -2 \sum_{\substack{i>j \\ \parallel \rm{NN}}} \left(J_{\parallel}\hat{S}^z_i \hat{S}^z_j + J_\parallel'[\hat{S}^x_i \hat{S}^x_j+\hat{S}^y_i \hat{S}^y_j]\right) \\
        &-2 \sum_{\substack{i>j \\ \perp \rm{NN}}} \left(J_{\perp} \hat{S}^z_i \hat{S}^z_j + J_\perp'[\hat{S}^x_i \hat{S}^x_j+\hat{S}^y_i \hat{S}^y_j]\right),\nonumber
\end{align}
where the summations are taken over nearest neighbours, NN, in-plane ($\perp \rm{NN}$) or out-of-plane ($\parallel \rm{NN}$). They concluded that transverse components of both exchange interactions are small $J_\perp'\approx\qty{-0.1}{\K}\times k_B\approx J_\parallel'$. The exchange interactions of $J_{\perp}$ and $J_{\parallel}$ are $J_{\perp}/k_B = -0.70$\,K and $J_{\parallel}/k_B = 0.07$\,K. Thus, the Hamiltonian is quasi two-dimensional Ising-like.
As calculated in Ref.\,\cite{Marti1995}, the dipolar energy per ion (33.85\,mK/ion) is much smaller than the N\'eel temperature and hence it was not included in Ref.\,\cite{Luis1998}.

\begin{figure}[tb]
    \includegraphics[width=\linewidth]{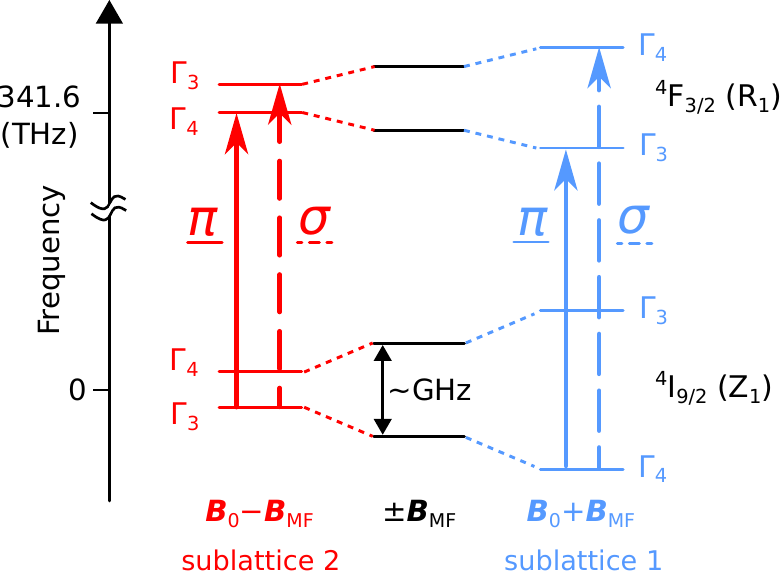}
    \caption{\label{fig: energy level diagram} Energy level diagram of the Nd\textsuperscript{3+} ions in an antiferromagnetic NdGaO\textsubscript{3} crystal. The red and blue arrows indicate allowed transitions of $\pi$ and $\sigma$ polarised light. Irreducible representations are from Ref.\,\cite{Koster_1963}. $\bm{B}_{0}$ is the applied static magnetic field and $\bm{B}$\textsubscript{MF} is the mean magnetic field. These magnetic fields are along the $c$ axis.}
\end{figure}

The optical absorptions we measured are from the $^4$I$_{9/2} (Z_1)$ to $^4$F$_{3/2} (R_1)$ doublets as shown in Fig.\,\ref{fig: energy level diagram}.
$Z_1$ is used as a label to represent the first (lowest) doublet among the $^4$I$_{9/2}$ states; similarly $R_1$ is used to represent the lowest doublet among the $^4$F$_{3/2}$ states.
The transition wavelength between these two doublets is near 878\,nm \cite{Orera1995_1, Orera1995_2}.

\subsection{\label{subsec: Hamiltonian} Model Hamiltonians}
We calculate energy levels of an antiferromagnetic NdGaO\textsubscript{3} crystal, based on an effective Hamiltonian of single rare-earth ions in crystals with no long-range magnetic ordering.
Rather than a rigorous treatment of magnetic interactions between rare-earth ions, we introduce a mean-field approach, where those magnetic interactions are incorporated as a \emph{mean} field in that effective Hamiltonian. A similar approach was adopted for antiferromagnetic ErLiF\textsubscript{4}, and the calculated energy levels showed good agreement with measured spectra \cite{Berrington_2022}.

The effective Hamiltonian of the single Nd\textsuperscript{3+} ion takes the following form
\begin{align}
\label{eq: conventional Hamilotnian}
    \hat{H}_\mathrm{Nd} = \hat{H}_{\text{FI}} + \hat{H}_{\text{CF}} + \hat{H}_{\text{Z}},
\end{align}
where $\hat{H}_{\text{FI}}$ is the free-ion interactions, such as the Coulomb interactions and the spin-orbit interactions. These interactions are spherically symmetric. The coefficients of the free-ion interactions are expected to be nearly independent of any crystal host or surrounding atoms \cite{Carnall_1989}. $\hat{H}_{\text{CF}}$ represents the crystal field interactions given as
\begin{align}
    \hat{H}_{\text{CF}} = \hspace{-0.6cm}\sum_{\substack{k=2,4,6\\q=-k,-k+2,\ldots,k}}\hspace{-0.6cm} B^{k}_{q} \hat{C}^{k}_{q}
\end{align}
where $\hat{C}^{k}_{q}$ is a spherical tensor operator with rank $k$ and component $q$. The corresponding $B^{k}_{q}$ is called the crystal field parameter. $\hat{H}_{\text{Z}}$ is the Zeeman interaction
\begin{align}
    \hat{H}_{\text{Z}} = -\bm{B}_0 \cdot \hat{\bm{\mu}},
\end{align}
where $\bm{B}_0$ is an applied magnetic field and $\hat{\bm{\mu}}$ is the magnetic moment of the Nd$^{3+}$ ion in the crystal defined as $\hat{\bm{\mu}} = -\mu_B(\hat{\bm{L}}+ 2\hat{\bm{S}})$ with total orbital and spin angular momenta.
By including a mean magnetic field from interactions between Nd$^{3+}$ ions, $\bm{B}_{\text{MF}}$, in $\hat{H}_{\text{Z}}$ such that $\bm{B}_0 \rightarrow \bm{B}_0 + \bm{B}_{\text{MF}}$, the effective Hamiltonian of Eq.\,\eqref{eq: conventional Hamilotnian} becomes
\begin{align}
\label{eq: effective Hamitonian}
    \hat{H}_\mathrm{Nd'} = \hat{H}_{\text{FI}} + \hat{H}_{\text{CF}} + \hat{H}_{\text{Z}'},
\end{align}
with
\begin{align}
    \hat{H}_{\text{Z}'} = -\left( \bm{B}_0 + \bm{B}_{\text{MF}} \right) \cdot \hat{\bm{\mu}}.
\end{align}

For the calculation, we used free-ion parameters for Nd:LaF\textsubscript{3} from Ref.\,\cite{Carnall_1989} and crystal field parameters $B^{k}_{q}$ for NdGaO\textsubscript{3} from Ref.\,\cite{Novak2013}. The operators of the Nd$^{3+}$ ion were from the \texttt{dieke} package in Python \cite{dieke}.
The mean field is calculated from the exchange interaction given in Ref.~\cite{Luis1998}. In the antiferromagnetic phase with an applied field along $c$, there are four antiferromagnetic bonds with strength $J_\perp$ and two ferromagnetic bonds with strength $J_\parallel$. The mean field is therefore 
\begin{equation}
    \bm{B}_\mathrm{MF} = 
        \frac{2J_\parallel \mp 4J_\perp}{\mu_B g_c}\bm{\hat{z}},
\end{equation}
where $g_c$ is the ground state $g$-factor along the $c$ axis, and the minus-sign is taken in the antiferromagnetic phase and positive in the paramagnetic phase.

We refer to selection rules to assign energy levels from observed spectra. The selection rules of electric dipole transitions between $^4$I$_{9/2} (Z_1)$ and $^4$F$_{3/2} (R_1)$ states can be determined using the effective Hamiltonian of Eq.\,\eqref{eq: effective Hamitonian} and their irreducible representations. On the $C_s$ site symmetry, both $^4$I$_{9/2} (Z_1)$ and $^4$F$_{3/2} (R_1)$ doublets are characterised by the summation of two irreducible representations, $\Gamma_3 + \Gamma_4$ (using the notation of Ref.\,\cite{Koster_1963}). The doublets split if the time reversal symmetry is broken by, for instance, applying a nonzero static magnetic field or the NdGaO\textsubscript{3} crystal being magnetically ordered. The resultant four states can be expressed by a superposition as $\alpha | \Gamma_3 \rangle + \beta | \Gamma_4 \rangle$ where $\alpha$ and $\beta$ are complex coefficients satisfying $|\alpha|^2 + |\beta|^2 = 1$. These coefficients generally depend on magnetic field strength and direction \cite{Guillot_2005, Afzelius_2010}.
We determined the irreducible representations by diagonalising the Hamiltonian of Eq.\,\eqref{eq: effective Hamitonian}. The electronic wavefunctions take the following form:
For $^4$I$_{9/2} \, (Z_1)$,
\begin{align}
\label{eq: 4I9/2 representation}
    \lambda_1 \lvert {^4}I_{9/2}, M_J &= \pm 1/2 \rangle + \lambda_2 \lvert {^4}I_{9/2}, M_J = \mp 7/2 \rangle \nonumber \\
    &\quad \quad \quad \quad \quad + \lambda_3 \lvert {^4}I_{9/2}, M_J = \pm 9/2 \rangle,
\end{align}
and for $^4$F$_{3/2} \, (R_1)$,
\begin{align}
\label{eq: 4F3/2 representation}
    \lambda_4 \lvert {^4}F_{3/2}, M_J = \pm 1/2 \rangle + \lambda_5 \lvert {^4}F_{3/2}, M_J = \mp 3/2 \rangle,
\end{align}
where $\lambda_i$ are complex coefficients.
In Eqs.\,\eqref{eq: 4I9/2 representation} and \eqref{eq: 4F3/2 representation}, each of the two expressions corresponds to a different irreducible representation, $\Gamma_3$ and $\Gamma_4$, respectively. The states with an $M_J=+1/2$ component are $\Gamma_3$. With the applied magnetic field along the $c$ axis, the calculation results indicate the four states, from the ground state, $\Gamma_4-\Gamma_3-\Gamma_3-\Gamma_4$ ($\Gamma_3-\Gamma_4-\Gamma_4-\Gamma_3$) for sublattice 1 (sublattice 2) as shown in Fig.\,\ref{fig: energy level diagram}. From the character table and their bases of the $C_s$ site symmetry in Ref.\,\cite{Koster_1963}, the electric dipole transition from the ground to the lower (higher) optically excited state can occur with electric field polarised along (perpendicular to) the $c$ axis. These two conditions are referred to as $\pi$ and $\sigma$ polarisations, respectively.

To describe some of the observed lines it was necessary to expand our model beyond a single Nd\textsuperscript{3+} ion. We used the following Hamiltonian to model a coupled pair of ions:
\begin{align}
\label{eq:pairham}
    \hat{H}_{\mathrm{Nd} \otimes \mathrm{Nd}} ={}& \hat{H}_{\text{Nd},1} + \hat{H}_{\text{Nd},2} -2\bm{\hat{\mu}}_1\cdot \bm{\tilde{J}}_{12} \cdot\bm{\hat{\mu}}_2 \nonumber \\& -\bm{B}_{\mathrm{MF},1}\cdot\bm{\hat{\mu}}_1 -\bm{B}_{\mathrm{MF},2}\cdot\bm{\hat{\mu}}_2,
\end{align}
with
\begin{align}
    \bm{B}_{\text{MF},i} ={}&2\sum_{\substack{j\in\perp \mathrm{NN}\\ j\neq1,2}}\langle\bm{\hat{\mu}}_j\rangle\cdot\bm{\tilde{J}}_\perp + 2\sum_{\substack{j\in\parallel \mathrm{NN}\\ j\neq1,2}}\langle\bm{\hat{\mu}}_j\rangle\cdot\bm{\tilde{J}}_\parallel.
\end{align}
The exchange coupling $\bm{\tilde{J}}_{12}$ and mean fields $\bm{B}_{\text{MF},1,2}$ depend on the pair of ions considered; the mean field counts the other ions not considered in the pair.
At \qty{0}{\kelvin}, the spins are stationary such that $\langle\bm{\hat{\mu}}_j\rangle=\pm \bm{\hat{z}}\mu_B g_c/2$.
For an in-plane pair $\bm{\tilde{J}}_{12} = \bm{\tilde{J}}_\perp$, and $\bm{B}_{\mathrm{MF},1} = \bm{\hat{z}}(2J_\parallel \mp 3J_\perp)/(\mu_B g_c) = \mp\bm{B}_{\mathrm{MF},2}$, whereas
for an out-of-plane pair $\bm{\tilde{J}}_{12} = \bm{\tilde{J}}_\parallel$ and $\bm{B}_{\mathrm{MF},1} = \bm{\hat{z}}(J_\parallel \mp 4J_\perp)/(\mu_B g_c) = \bm{B}_{\mathrm{MF},2}$. 
We took the exchange-coupling tensors to be
\begin{equation}
    \bm{\tilde{J}}_{\parallel,\perp} = \frac{1}{\mu_B^2}\mathop{\mathrm{diag}}\left(\frac{J_{\parallel,\perp}'}{g_a^2},\frac{J_{\parallel,\perp}'}{g_b^2},\frac{J_{\parallel,\perp}}{g_c^2}\right),
\end{equation}
where $g_{a,b,c}$ are the $g$-factor along the $a,b,c$ axes. Here we made a simplifying assumption, choosing to make the exchange tensors diagonal in the crystallographic axes, and opting to take the same crystal-field parameters for all ions, though they and their $g$-tensors are oriented differently in the $ab$ plane. Because $J_\perp',J_\parallel'$ are small and we will be examining regions of constant magnetization, these assumptions are expected to only have a small effect on our results in that plane. 
The factors of $1/(\mu_B^2g^2)$ were necessary to convert between an effective-spin-1/2 model [Eq.\,\eqref{eq: Luis Hamiltonian}] and our magnetic-moment model, and ensure the exchange energies are the same.
The model in Eq.\,\eqref{eq:pairham} allows calculation of simultaneous excitation of two magnetically coupled Nd\textsuperscript{3+} ions by a single photon.

The selection rules for the pair transitions can be derived from the single-ion transitions. An antiferromagnetic pair has one each of its ions in the $\Gamma_3$, $\Gamma_4$ irreducible representations. The pair is therefore $\Gamma_3\Gamma_4=\Gamma_1$. By examining the pair's irreducible representations in this way for the ground and excited states, we determined pair selection rules; a transition between the ground Kramers doublet of the second ion swaps the polarisation required to excite the transition. The selection rules are given in Table~\ref{tab:selectionrules}.
They are expected to hold well for in-plane pairs as those ions share a common mirror plane, but may not hold for out-of-plane pairs as excitation of one ion breaks the mirror symmetry at the other.

\begin{table}
    \caption{Selection rules for single- and two-ion transitions in $C_s$ symmetry for given initial and final irreducible representation (irrep.) \cite{Koster_1963}. 
    $\sigma\rightarrow E_x,E_y,B_z$; $\pi\rightarrow B_x,B_y,E_z$. Our transition is not magnetic-dipole allowed, but the magnetic-field components have been included for reference.
    }
    \label{tab:selectionrules}
    \centering
    \begin{tabular}{c|c|c}
        initial irrep. & final irrep. & polarization  \\ \hline \hline
        $\Gamma_3$ & $\Gamma_3$ & $\sigma$ ($\Gamma_1$) \\
        $\Gamma_4$ & $\Gamma_4$ & $\sigma$ ($\Gamma_1$) \\
        $\Gamma_3$ & $\Gamma_4$ & $\pi$ ($\Gamma_2$)\\
        $\Gamma_4$ & $\Gamma_3$ & $\pi$ ($\Gamma_2$)\\
        \hline
        $\Gamma_1=\Gamma_3\Gamma_4=\Gamma_4\Gamma_3$ & $\Gamma_1=\Gamma_3\Gamma_4=\Gamma_4\Gamma_3$ & $\sigma$ ($\Gamma_1$) \\
        $\Gamma_2=\Gamma_3\Gamma_3=\Gamma_4\Gamma_4$ & $\Gamma_2=\Gamma_3\Gamma_3=\Gamma_4\Gamma_4$ & $\sigma$ ($\Gamma_1$) \\
        $\Gamma_1=\Gamma_3\Gamma_4=\Gamma_4\Gamma_3$ & $\Gamma_2=\Gamma_3\Gamma_3=\Gamma_4\Gamma_4$ & $\pi$ ($\Gamma_2$)\\
        $\Gamma_2=\Gamma_3\Gamma_3=\Gamma_4\Gamma_4$ & $\Gamma_1=\Gamma_3\Gamma_4=\Gamma_4\Gamma_3$ & $\pi$ ($\Gamma_2$)\\
    \end{tabular}
\end{table}

\subsection{\label{subsec: Sample preparation and experimental technique}Sample preparation and experimental technique}
\begin{figure*}[tb]
    \begin{center}
        \includegraphics[width=0.9\linewidth]{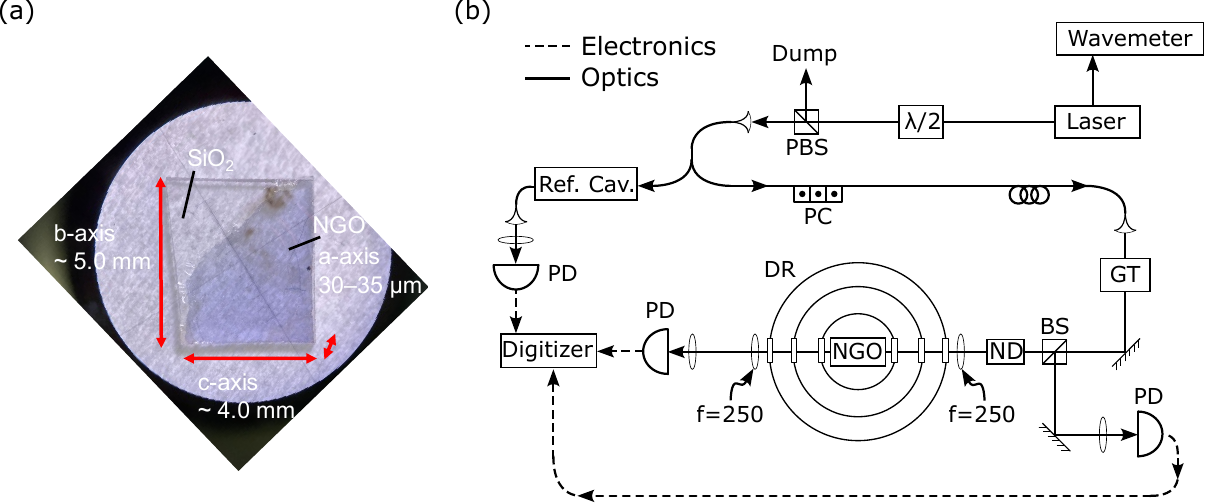}
        \caption{\label{fig: MW and Opt setup} (a) A photograph of the thinned down NdGaO$_3$ sample attached on a quartz wafer substrate used for optical measurements.
        (b) Setup of optical measurements. Solid and dashed arrows represent, respectively, optical and electric lines. PBS denotes Polarised Beam Splitter cube; ND, neutral density filter; BS, unpolarised Beam Splitter cube; PC, fibre Polarisation Controller; Ref.\,Cav.\,, Reference Cavity; GT, Glan–Taylor polariser; DR, Dilution Refrigerator; PD, Photodetector; NGO, NdGaO$_3$ sample; $\lambda/2$, half-wave plate.}
    \end{center}
\end{figure*}

The NdGaO$_3$ crystal used for this study was purchased from MSE Supplies LLC (Tucson, Arizona, USA). It was initially $(0.5, 5.0, 5.0)$\,mm in size along the ($a$, $b$, $c$) axes.
The sample was attached by epoxy onto a quartz wafer substrate with thickness of 0.5\,mm. Quartz was chosen as it is transparent in our optical wavelength range and has high thermal conductivity at dilution refrigerator temperatures \cite{Ekin_2006}.
Then, the sample was polished down to a thickness of approximately 30 to \qty{35}{\micro\meter} thick, as measured by a micrometer.
Figure \ref{fig: MW and Opt setup}(a) shows the thinned-down sample on the substrate, showing the uneven shape resulting from polishing.

The sample was mounted in a copper holder on the end of a cold finger attached to the cold plate of a dilution refrigerator. The copper holder was also a microwave loop-gap resonator \cite{hiraishi_long_2025}, although no microwave measurements were made here. A sensor on the cold finger indicated a base temperature of around 40\,mK. The copper holder was mounted in the bore of a 3\,T superconducting magnet, such that there was free-space optical access to the sample, with the beam path orthogonal to the magnetic field. Two different sets of measurements were made: one with the magnetic field parallel to the crystal $c$ axis, and one with the magnetic field parallel to the $b$ axis. The sample was rotated in the holder to make these different measurements.

Figure \ref{fig: MW and Opt setup}(b) shows the configuration for optical measurements. We used a titanium-doped sapphire (Ti:S) laser. 
The Ti:S laser could sweep its frequency over 30\,GHz at a time. The full spectra were made by stitching these \qty{30}{\GHz} chunks together. To enable the frequency axes of each of these sweeps to be calibrated, a reading from a wavemeter (HighFinesse WS/7) was taken at the start and the end of each scan and the transmission of a fibre ring resonator was monitored as the data was taken. The fibre resonator was made by splicing two of the fibres on a fibre splitter together and the different longitudinal modes provided a fine frequency ruler (free spectral range of 149 MHz). The fibre loop resonator was placed in an aluminium die-cast box full of sand, which in turn was insulated with polystyrene and placed in a wooden box. The temperature of the die cast box was temperature stabilised to a few mK, a few degrees above room temperature. This reduced any thermal drift to a level well below what would affect these results.

Optical polarisation was adjusted with a manual fibre polarisation controller and a Glan-Taylor polariser. Optical input power was adjusted by neutral density filters. Two photodetectors were used: a photomultiplier tube (PMT, Hamamatsu Photonics R1477-06) and a silicon photodiode (Si PD). When measuring with PMT, the power into the fridge was set to 80\,nW, and the output was fibre-coupled, such that the PMT could be far from the applied magnetic field. When measuring with the Si PD, the power into the fridge was set to \qty{1.1}{\micro\watt}. These powers were a compromise between signal-to-noise ratio and sample heating.
We swept optical wavelengths at a fixed magnetic field and measured optical transmission through the NdGaO$_3$ sample. This was repeated with various magnetic fields.

\section{Results and Discussion}
\begin{center}
\textbf{$\bm{B}_0$ $\parallel$ $c$ axis}    
\end{center}

\begin{figure*}[!tb]
\includegraphics[width=\linewidth]{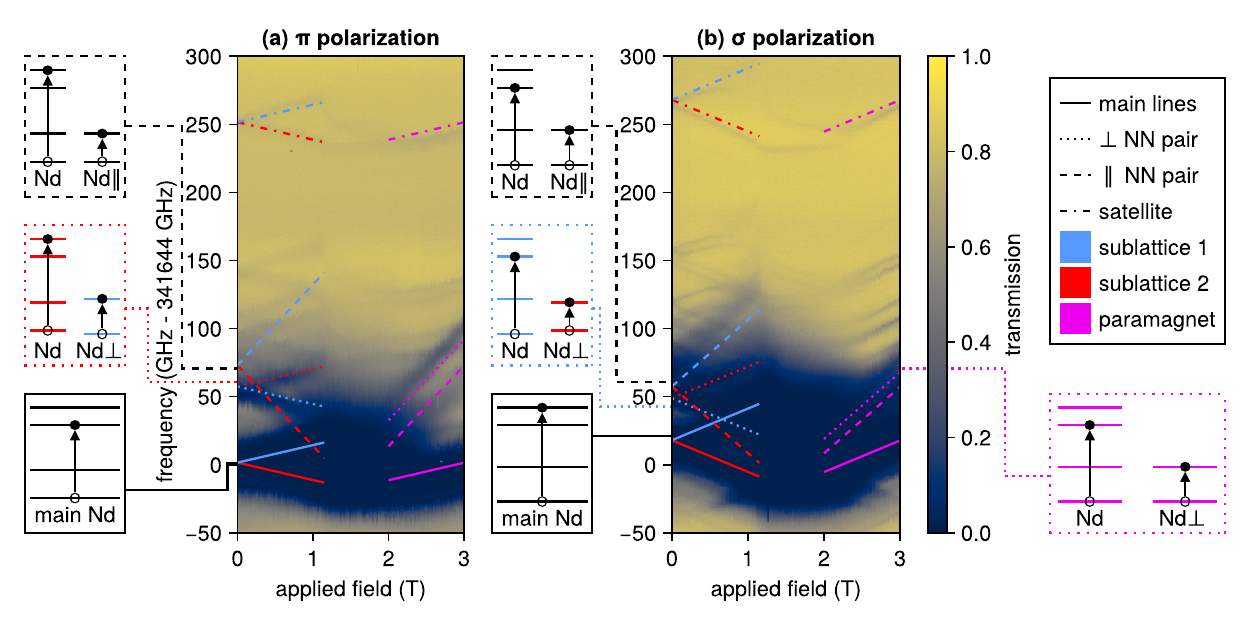}
\caption{\label{Fig: Opt B parallel c-axis} Optical transmission spectra with the external static magnetic fields along the $c$ axis and the (a) $\pi$- and (b) $\sigma$-polarised electric field. Solid, dotted, and dashed lines represent main, in-plane-pair ($\perp$ NN), and out-of-plane-pair ($\parallel$ NN) lines, respectively, which are calculated using the Hamiltonian in Eq.\,\eqref{eq:pairham}, with $J_\perp/k_B = \qty{-0.65}{\kelvin}$, $J_\parallel/k_B=+\qty{0.07}{\kelvin}$, and $J_\parallel'=J_\perp'=\qty{-0.1}{\kelvin}\times k_B$.
Calculated transition frequencies are shown, with energy level diagrams showing their physical mechanisms, for the polarisation that the calculated selection rules allow.
There is a global frequency offset for all lines as the crystal-field model, which was not trained on any optical data, gives a different transition frequency. An example set of satellite lines are shown dash-dotted, and these are simply calculated as main lines, frequency offset by \qty{250}{\GHz}.  All spectra were measured by the PMT at the base temperature of the DR, below 40\,mK.}
\end{figure*}

Figure \ref{Fig: Opt B parallel c-axis} shows optical transmission spectra under a magnetic field up to 3 T along the $c$ axis with $\pi$ and $\sigma$ polarised electric field.
We observed absorptions with varying intensities, ranging from barely resolvable to strong, saturated features.

The strongest absorption features are single-Nd transitions, which we label `main lines.'
At zero magnetic field, $\pi$- and $\sigma$-polarised light induces different optical transitions, consistent with the optical selection rules predicted in Sec.\,\ref{subsec: Hamiltonian}. 
As predicted in Fig.\,\ref{fig: energy level diagram} the $\pi$ polarised light is absorbed at a lower frequency, to the lower of the excited states, while the $\sigma$ polarised is absorbed at higher frequency, to the upper excited state.
These absorption features occur at about \qtylist{0;20}{\GHz} in Fig.~\ref{Fig: Opt B parallel c-axis}. The energy separation between the $R_1$ doublet is due to the mean field of the magnetic interactions between the antiferromagnetically ordered Nd\textsuperscript{3+} ions. 
An external static magnetic field differentiates optical transition frequencies between the two sublattices: electronic magnetic moments are parallel to the applied magnetic field at one sublattice, and anti-parallel at the other. The optical transition frequencies are linearly proportional to a small applied magnetic field due to the Zeeman effect, which shows that the magnetic structure is not changing.

The optical spectra have bends at about 1.1\,T and 2.3\,T, suggesting that three magnetic phases are present. We shall call these phases antiferromagnetic (0--1.1\,T), intermediate (1.1--2.3\,T), and paramagnetic (above 2.3\,T). 
We have calculated transition frequencies in the antiferromagnetic and paramagnetic phases because there the magnetisation does not vary with field. In the paramagnetic phase, the sublattices become equivalent, and so there are half as many lines as in the antiferromagnetic phase.

In addition to the main lines, we see other features, which we call `two-Nd lines' and `satellite lines,' in terms of their mechanisms.
The two-Nd lines are due to simultaneous excitations of adjacent two Nd\textsuperscript{3+} ions by a single photon. The two-Nd lines were at higher frequencies than those of main lines; about $+50$\,GHz at 0\,T (Fig.\,\ref{Fig: Opt B parallel c-axis}). They are also saturated at some points, but are weaker than the main lines. Absorption frequencies are equal to the sum of each of the two Nd\textsuperscript{3+} ions' absorption frequencies with an additional energy difference due to the exchange coupling between the ion pair, resulting in higher frequencies than those of the main lines.
The slopes of these two-Nd lines provide evidence that there is both ferro- and antiferromagnetic ordering in the crystal.

We observed all but one of the optical two-Nd excitations predicted by our model. The selection rules for these composite transitions agree with Tab.~\ref{tab:selectionrules}. 
The transition frequencies were calculated using Hamiltonian in Eq.\,\eqref{eq:pairham}, in which we took exchange parameters from Ref.~\cite{Luis1998}, but altered one value; they determined that $J_\perp/k_B = \qty{-0.7}{\kelvin}$, whereas we took $J_\perp/k_B = \qty{-0.65}{\kelvin}$.
We would like to emphasise here that all parameters were taken directly from the literature, except for this one modification.
The $\pi$-polarized line where an Nd\textsuperscript{3+} ion and its out-of-plane nearest neighbour are excited was not observed, but we see in Fig.\,\ref{Fig: Opt B parallel c-axis} that $\pi$-polarized transitions are weaker than $\sigma$, and the exchange coupling is weaker between this pair.

A number of weaker absorption features, which we call satellite lines, were observed at several frequencies ranging from $-50$ to $+250$\,GHz away from the main line. One set, \qty{250}{\GHz} higher in frequency than the main line has been marked on Fig.\,\ref{Fig: Opt B parallel c-axis}.
From the literature, the satellite lines have been considered to originate from a rare-earth ion adjacent to an impurity or other defect in the crystal. This creates a distortion of the crystal lattice and results in the active ion's optical transition frequencies being outside of the inhomogeneous broadening \cite{Cone_1993, Jaaniso1994, Yamaguchi1999, Ahlefeldt2013}. Similar satellite lines have been observed in various stoichiometric rare-earth crystals such as EuCl$_3 \cdot$6H$_2$O \cite{Ahlefeldt2013} and ErLiF$_4$ \cite{Berrington_2022}. 

Figure \ref{fig: B parallel c-axis, satellite line} show an enlarged section of Fig.\,\ref{Fig: Opt B parallel c-axis}.
\begin{figure*}
\includegraphics[width=0.9\linewidth]{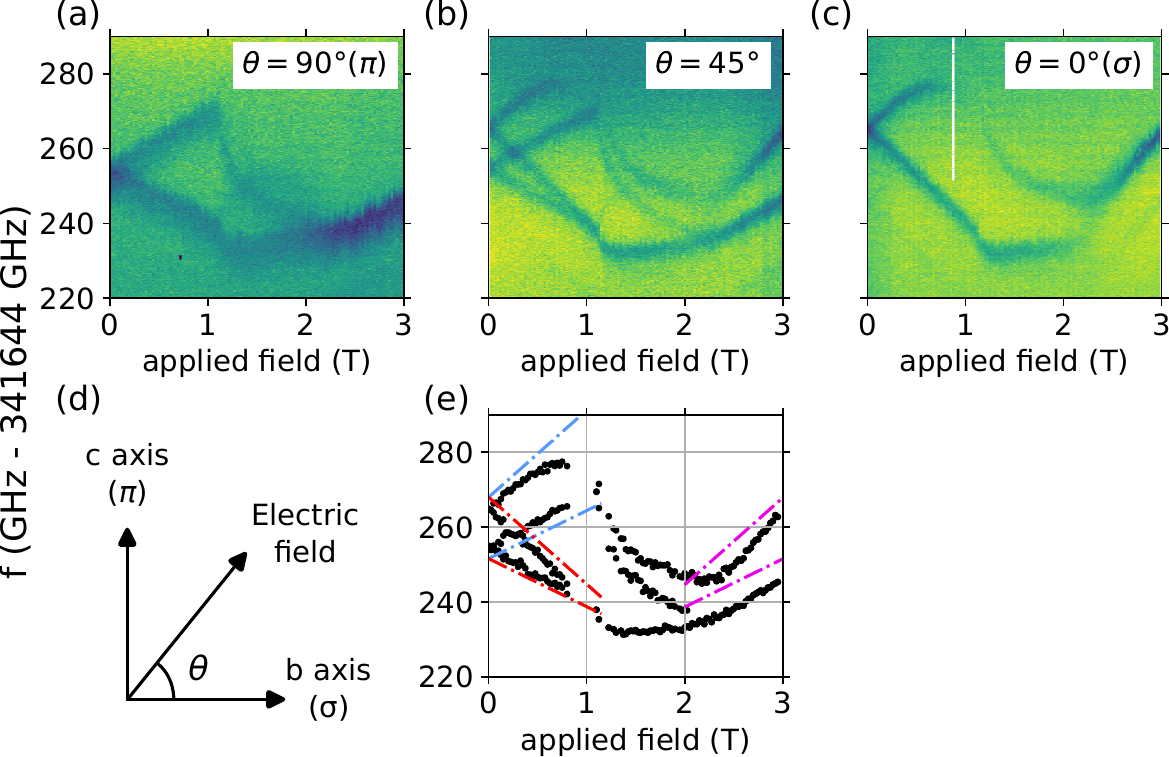}
\caption{\label{fig: B parallel c-axis, satellite line} Magnetic field dependence of the satellite line spectra with (a) $\pi$, (b) $45^\circ$, and (c) $\sigma$ optical polarisations. Panels (a) and (c) are details from Fig.\,\ref{Fig: Opt B parallel c-axis}. In panel (c), a white vertical line at 0.89\,T represents an experimental error that is irrelevant to the absorption. (d) A diagram of an optical polarisation angle and the crystal axes.
(e) Black solid dots show extracted absorption frequencies from panels (a) and (c), with the same lines as in Fig.\,\ref{Fig: Opt B parallel c-axis}. }
\end{figure*}
Unlike the main lines, the absorption was unsaturated and hence resolved. In the antiferromagnetic phase ranging from 0 to 1.1\,T, the spins remain in the $c_z$ configuration, which does not vary energy by exchange and magnetic dipole-dipole interaction. Hence, the magnetic field dependence of the transition frequency is purely due to the Zeeman interaction, which is proportional to the external magnetic field strength.
Near the phase transition, above about \qty{0.8}{\tesla}, sublattice 1 (blue) shows a curved, weak transition in the $\sigma$ polarisation. This is presumably due to a changing mean field. 
The mean field comes largely from sublattice 2 (red), which has transition frequencies that become nearly equal, meaning that its mean-field nearly cancels the applied field. Hence, there may be thermal population of the upper Zeeman level of sublattice 2, which alters the exchange interaction to sublattice 1.

It is common in antiferromagnetic systems for a ``spin-flop" phase to exist between the antiferromagnetic and paramagnetic phases. In a classical spin-flop state, all spins are at the same angle from the $c$ axis. Therefore, we would expect to see lines from a single sublattice. In our intermediate phase, we see enough lines for two sublattices, suggesting more complicated behaviour.

In the intermediate phase the transitions vary non-linearly with applied magnetic field and join continuously with the transitions of the paramagnetic phase. This suggests that in our intermediate phase, like the classical spin-flop phase, the spins continuously rotate as the field is increased to reach the paramagnetic phase in a type II phase transition.
Developing a quantitative model for the optical spectra in this phase remains work in progress.
Such nonlinear field-dependence has also been seen in ErLiF$_4$, another rare-earth antiferromagnetic material \cite{Berrington_2022}.

In the paramagnetic phase (above 2.3\,T), the measured absorption frequencies increase almost linearly. In both $\pi$ and $\sigma$ polarised spectra, there is a single absorption feature, compared to two in the antiferromagnetic phase.
In the paramagnetic phase, the linear dependence implies the Zeeman interaction is dominant compared with other magnetic interactions. There, the relative spins vary little. The single absorption feature implies that the two sublattices become identical above 2.3\,T.

Figure \ref{fig: B parallel c-axis, satellite line}(e) shows calculated transition frequencies plotted on the measured spectra. In both antiferromagnetic and paramagnetic phases, the calculated spectra are in reasonable agreement with the measured spectra.

\begin{center}
    \textbf{$\bm{B}_0$ $\parallel$ $b$ axis}
\end{center}

In the same way, we also measured absorption spectra with the field along the $b$ axis. For these measurements, in addition to using the PMT for $\pi$-polarisation measurements, a silicon photodiode was used for the $\sigma$ polarisation, such that we could make measurements with higher optical power.
\begin{figure*}[!tb]
\includegraphics[width=\linewidth]{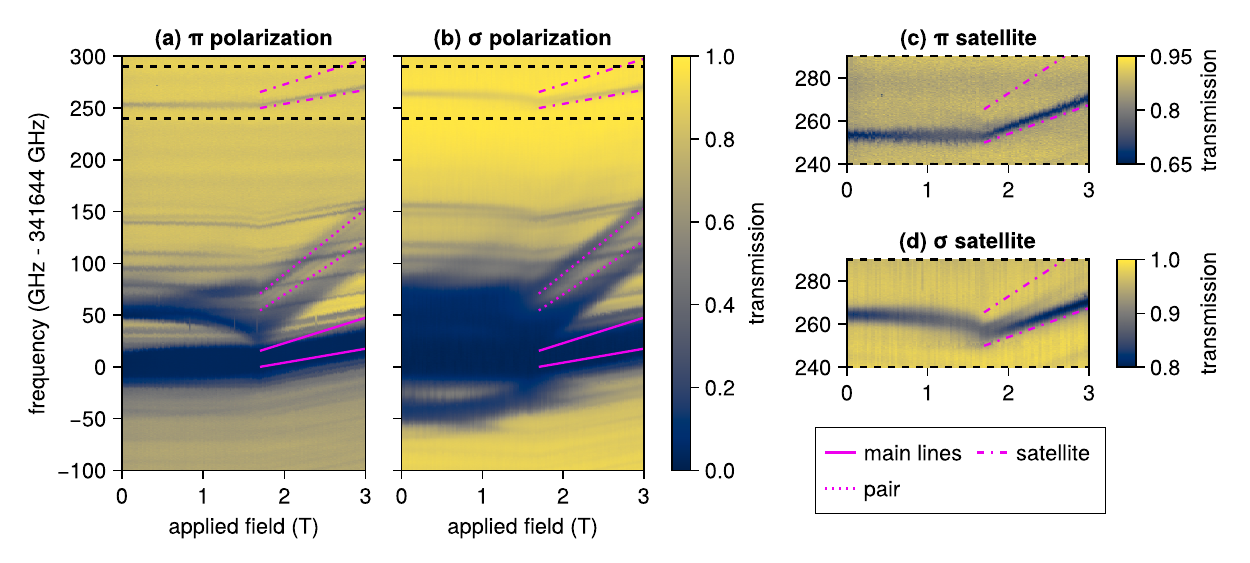}
\caption{\label{Fig: Opt B parallel b-axis} Optical transmission spectra with the external static magnetic fields along the $b$ axis. (a) $\pi$-polarised spectra measured with the PMT.
(b) $\sigma$-polarised spectra measured with the Si PD. 
Calculated transition frequencies are shown for the paramagnetic phase.
(c,d) Zoomed-in view on a different colour scale of the examined satellite line shown as dashed lines in (a,b), showing the two main absorption lines.
Both spectra were measured at the base temperature of the DR, below 40\,mK.}
\end{figure*}

Figure \ref{Fig: Opt B parallel b-axis} shows the entire spectra including the main, two-Nd, and satellite lines with two optical polarisations. In addition to the main, two-Nd, and satellite lines, another absorption was observed only with the $\sigma$ polarised light and higher optical input power located at about \qty{-30}{\GHz}. This absorption occurs as a result of a thermal excitation from the ground state. The absorption frequency is lower than that of the main line and determined by the difference between the frequencies of the main line and the first excited state.

In the measured field range, one magnetic phase transition was observed at around 1.72\,T, where a discrete change of the magnetic field dependence of the absorption frequencies was observed. This field is consistent with the phase transition from the antiferromagnetic to the paramagnetic phases \cite{blaauw-smith_magnetostriction_2024}. 
Above this field, all the spins are aligned ferromagnetically along the external magnetic field direction.

In the paramagnetic phase, the mean field due to the exchange interaction is oriented along the crystallographic $b$ axis. We model the transitions in this phase with the Hamiltonian of Eq.\,\eqref{eq:pairham}. Because $J_\perp'\approx J_\parallel'$, for both kinds of pairs, $\bm{B}_{\mathrm{MF},1}=5J_\perp'\bm{\hat{y}}/(g_b\mu_B)=\bm{B}_{\mathrm{MF},2}$.
With this applied field, the eigenstates mix $\Gamma_3$ and $\Gamma_4$, hence the selection rules are broken.
Both predicted two-ion excitations are seen readily on the spectra of both polarisations.

We discuss the optical transition properties, with a particular focus on the satellite lines. Figure \ref{Fig: Opt B parallel b-axis}(c,d) shows zoomed spectra of the \qty{250}{\GHz} offset satellite lines. In the antiferromagnetic phase up to 1.72\,T, each $\pi$- and $\sigma$-polarised condition showed only a single absorption unlike those with $\bm{B}_0 \parallel c$ axis (Fig.\,\ref{fig: B parallel c-axis, satellite line}). The single absorption is due to the magnetic field direction being normal to the easy axis, which makes the resonances identical for each sublattice, even in the antiferromagnetic phase.

In the paramagnetic phase above 1.72\,T, we see a main line in both polarizations due to the broken selection rules. The second main line is weakly seen in the $\sigma$ polarisation.

\section{Conclusions}
We have performed high-resolution optical spectroscopy on an antiferromagnetic NdGaO\textsubscript{3} crystal.
Three magnetic phases are present in the measured magnetic field range, 0--3\,T.
The intermediate phase between the antiferromagnetic and paramagnetic phases is presumably distinct from a standard spin-flop phase.
The number of optical absorptions in this phase is inconsistent with that expected in the standard spin-flop phase, where two magnetic sublattices have a common canting angle from the applied field.
Satellite lines were observed, as seen in other stoichiometric rare-earth crystals.
The effective Hamiltonian modelled in this study explains well the measured absorptions: single-Nd and two-Nd transitions, and their selection rules determined by the irreducible representations of each Nd\textsuperscript{3+} electronic state.
The two-Nd model predicts simultaneous excitations of two adjacent magnetically coupled rare-earth ions.
We believe that this model is applicable to other stoichiometric rare-earth magnetic crystals.

This work opens up possibilities for exploring other stoichiometric rare-earth magnetic crystals, as it provides a method for assigning transitions and calculating their energies.

\section*{Acknowledgements}
The authors would like to thank Brent Pooley for thinning the sample, and David J. Prior for orienting its crystal axes.
Figs.~\ref{Fig: Opt B parallel c-axis} and \ref{Fig: Opt B parallel b-axis} were made using \texttt{Makie.jl} \cite{DanischKrumbiegel2021}.

This work was supported by Quantum Technologies Aotearoa, a research programme of Te Whai Ao -- the Dodd Walls Centre, funded by the New Zealand Ministry of Business Innovation and Employment through International Science Partnerships, contract number UOO2347.

\providecommand{\noopsort}[1]{}\providecommand{\singleletter}[1]{#1}%
%

% \bibliography{apssamp}% Produces the bibliography via BibTeX.

\begin{thebibliography}{43}%
\makeatletter
\providecommand \@ifxundefined [1]{%
 \@ifx{#1\undefined}
}%
\providecommand \@ifnum [1]{%
 \ifnum #1\expandafter \@firstoftwo
 \else \expandafter \@secondoftwo
 \fi
}%
\providecommand \@ifx [1]{%
 \ifx #1\expandafter \@firstoftwo
 \else \expandafter \@secondoftwo
 \fi
}%
\providecommand \natexlab [1]{#1}%
\providecommand \enquote  [1]{``#1''}%
\providecommand \bibnamefont  [1]{#1}%
\providecommand \bibfnamefont [1]{#1}%
\providecommand \citenamefont [1]{#1}%
\providecommand \href@noop [0]{\@secondoftwo}%
\providecommand \href [0]{\begingroup \@sanitize@url \@href}%
\providecommand \@href[1]{\@@startlink{#1}\@@href}%
\providecommand \@@href[1]{\endgroup#1\@@endlink}%
\providecommand \@sanitize@url [0]{\catcode `\\12\catcode `\$12\catcode `\&12\catcode `\#12\catcode `\^12\catcode `\_12\catcode `\%12\relax}%
\providecommand \@@startlink[1]{}%
\providecommand \@@endlink[0]{}%
\providecommand \url  [0]{\begingroup\@sanitize@url \@url }%
\providecommand \@url [1]{\endgroup\@href {#1}{\urlprefix }}%
\providecommand \urlprefix  [0]{URL }%
\providecommand \Eprint [0]{\href }%
\providecommand \doibase [0]{https://doi.org/}%
\providecommand \selectlanguage [0]{\@gobble}%
\providecommand \bibinfo  [0]{\@secondoftwo}%
\providecommand \bibfield  [0]{\@secondoftwo}%
\providecommand \translation [1]{[#1]}%
\providecommand \BibitemOpen [0]{}%
\providecommand \bibitemStop [0]{}%
\providecommand \bibitemNoStop [0]{.\EOS\space}%
\providecommand \EOS [0]{\spacefactor3000\relax}%
\providecommand \BibitemShut  [1]{\csname bibitem#1\endcsname}%
\let\auto@bib@innerbib\@empty
%</preamble>
\bibitem [{\citenamefont {Thiel}\ \emph {et~al.}(2011)\citenamefont {Thiel}, \citenamefont {B\"ottger},\ and\ \citenamefont {Cone}}]{thiel_rare-earth-doped_2011}%
  \BibitemOpen
  \bibfield  {author} {\bibinfo {author} {\bibfnamefont {C.~W.}\ \bibnamefont {Thiel}}, \bibinfo {author} {\bibfnamefont {T.}~\bibnamefont {B\"ottger}},\ and\ \bibinfo {author} {\bibfnamefont {R.~L.}\ \bibnamefont {Cone}},\ }\bibfield  {title} {\bibinfo {title} {Rare-earth-doped materials for applications in quantum information storage and signal processing},\ }\href {https://doi.org/10.1016/j.jlumin.2010.12.015} {\bibfield  {journal} {\bibinfo  {journal} {Journal of Luminescence}\ }\bibinfo {series} {Selected papers from {DPC}'10},\ \textbf {\bibinfo {volume} {131}},\ \bibinfo {pages} {353} (\bibinfo {year} {2011})}\BibitemShut {NoStop}%
\bibitem [{\citenamefont {Macfarlane}\ \emph {et~al.}(1998)\citenamefont {Macfarlane}, \citenamefont {Meltzer},\ and\ \citenamefont {Malkin}}]{macfarlane_optical_1998}%
  \BibitemOpen
  \bibfield  {author} {\bibinfo {author} {\bibfnamefont {R.~M.}\ \bibnamefont {Macfarlane}}, \bibinfo {author} {\bibfnamefont {R.~S.}\ \bibnamefont {Meltzer}},\ and\ \bibinfo {author} {\bibfnamefont {B.~Z.}\ \bibnamefont {Malkin}},\ }\bibfield  {title} {\bibinfo {title} {Optical measurement of the isotope shifts and hyperfine and superhyperfine interactions of {Nd} in the solid state},\ }\href {https://doi.org/10.1103/PhysRevB.58.5692} {\bibfield  {journal} {\bibinfo  {journal} {Physical Review B}\ }\textbf {\bibinfo {volume} {58}},\ \bibinfo {pages} {5692} (\bibinfo {year} {1998})}\BibitemShut {NoStop}%
\bibitem [{\citenamefont {Fraval}\ \emph {et~al.}(2004)\citenamefont {Fraval}, \citenamefont {Sellars},\ and\ \citenamefont {Longdell}}]{Fraval_2004}%
  \BibitemOpen
  \bibfield  {author} {\bibinfo {author} {\bibfnamefont {E.}~\bibnamefont {Fraval}}, \bibinfo {author} {\bibfnamefont {M.~J.}\ \bibnamefont {Sellars}},\ and\ \bibinfo {author} {\bibfnamefont {J.~J.}\ \bibnamefont {Longdell}},\ }\bibfield  {title} {\bibinfo {title} {Method of extending hyperfine coherence times in {Pr}\textsuperscript{3+}:{Y}\textsubscript{2}{Si}{O}\textsubscript{5}},\ }\href {https://doi.org/10.1103/PhysRevLett.92.077601} {\bibfield  {journal} {\bibinfo  {journal} {Physical Review Letters}\ }\textbf {\bibinfo {volume} {92}},\ \bibinfo {pages} {077601} (\bibinfo {year} {2004})}\BibitemShut {NoStop}%
\bibitem [{\citenamefont {Fraval}\ \emph {et~al.}(2005)\citenamefont {Fraval}, \citenamefont {Sellars},\ and\ \citenamefont {Longdell}}]{Fraval_2005}%
  \BibitemOpen
  \bibfield  {author} {\bibinfo {author} {\bibfnamefont {E.}~\bibnamefont {Fraval}}, \bibinfo {author} {\bibfnamefont {M.~J.}\ \bibnamefont {Sellars}},\ and\ \bibinfo {author} {\bibfnamefont {J.~J.}\ \bibnamefont {Longdell}},\ }\bibfield  {title} {\bibinfo {title} {Dynamic decoherence control of a solid-state nuclear-quadrupole qubit},\ }\href {https://doi.org/10.1103/PhysRevLett.95.030506} {\bibfield  {journal} {\bibinfo  {journal} {Physical review letters}\ }\textbf {\bibinfo {volume} {95}},\ \bibinfo {pages} {030506} (\bibinfo {year} {2005})}\BibitemShut {NoStop}%
\bibitem [{\citenamefont {B\"ottger}\ \emph {et~al.}(2009)\citenamefont {B\"ottger}, \citenamefont {Thiel}, \citenamefont {Cone},\ and\ \citenamefont {Sun}}]{Bottger_2009}%
  \BibitemOpen
  \bibfield  {author} {\bibinfo {author} {\bibfnamefont {T.}~\bibnamefont {B\"ottger}}, \bibinfo {author} {\bibfnamefont {C.~W.}\ \bibnamefont {Thiel}}, \bibinfo {author} {\bibfnamefont {R.~L.}\ \bibnamefont {Cone}},\ and\ \bibinfo {author} {\bibfnamefont {Y.}~\bibnamefont {Sun}},\ }\bibfield  {title} {\bibinfo {title} {Effects of magnetic field orientation on optical decoherence in {Er}\textsuperscript{3+}:{Y}\textsubscript{2}{Si}{O}\textsubscript{5}},\ }\href {https://doi.org/10.1103/PhysRevB.79.115104} {\bibfield  {journal} {\bibinfo  {journal} {Physical Review B}\ }\textbf {\bibinfo {volume} {79}},\ \bibinfo {pages} {115104} (\bibinfo {year} {2009})}\BibitemShut {NoStop}%
\bibitem [{\citenamefont {Lovri\ifmmode~\acute{c}\else \'{c}\fi{}}\ \emph {et~al.}(2011)\citenamefont {Lovri\ifmmode~\acute{c}\else \'{c}\fi{}}, \citenamefont {Glasenapp}, \citenamefont {Suter}, \citenamefont {Tumino}, \citenamefont {Ferrier}, \citenamefont {Goldner}, \citenamefont {Sabooni}, \citenamefont {Rippe},\ and\ \citenamefont {Kr\"oll}}]{Lovric_2011}%
  \BibitemOpen
  \bibfield  {author} {\bibinfo {author} {\bibfnamefont {M.}~\bibnamefont {Lovri\ifmmode~\acute{c}\else \'{c}\fi{}}}, \bibinfo {author} {\bibfnamefont {P.}~\bibnamefont {Glasenapp}}, \bibinfo {author} {\bibfnamefont {D.}~\bibnamefont {Suter}}, \bibinfo {author} {\bibfnamefont {B.}~\bibnamefont {Tumino}}, \bibinfo {author} {\bibfnamefont {A.}~\bibnamefont {Ferrier}}, \bibinfo {author} {\bibfnamefont {P.}~\bibnamefont {Goldner}}, \bibinfo {author} {\bibfnamefont {M.}~\bibnamefont {Sabooni}}, \bibinfo {author} {\bibfnamefont {L.}~\bibnamefont {Rippe}},\ and\ \bibinfo {author} {\bibfnamefont {S.}~\bibnamefont {Kr\"oll}},\ }\bibfield  {title} {\bibinfo {title} {Hyperfine characterization and spin coherence lifetime extension in {Pr}\textsuperscript{3+}:{La}\textsubscript{2}({WO}\textsubscript{4})\textsubscript{3}},\ }\href {https://doi.org/10.1103/PhysRevB.84.104417} {\bibfield  {journal} {\bibinfo  {journal} {Physical Review B}\ }\textbf {\bibinfo {volume} {84}},\ \bibinfo {pages} {104417} (\bibinfo {year}
  {2011})}\BibitemShut {NoStop}%
\bibitem [{\citenamefont {Zhong}\ \emph {et~al.}(2015)\citenamefont {Zhong}, \citenamefont {Hedges}, \citenamefont {Ahlefeldt}, \citenamefont {Bartholomew}, \citenamefont {Beavan}, \citenamefont {Wittig}, \citenamefont {Longdell},\ and\ \citenamefont {Sellars}}]{Zhong_2015}%
  \BibitemOpen
  \bibfield  {author} {\bibinfo {author} {\bibfnamefont {M.}~\bibnamefont {Zhong}}, \bibinfo {author} {\bibfnamefont {M.~P.}\ \bibnamefont {Hedges}}, \bibinfo {author} {\bibfnamefont {R.~L.}\ \bibnamefont {Ahlefeldt}}, \bibinfo {author} {\bibfnamefont {J.~G.}\ \bibnamefont {Bartholomew}}, \bibinfo {author} {\bibfnamefont {S.~E.}\ \bibnamefont {Beavan}}, \bibinfo {author} {\bibfnamefont {S.~M.}\ \bibnamefont {Wittig}}, \bibinfo {author} {\bibfnamefont {J.~J.}\ \bibnamefont {Longdell}},\ and\ \bibinfo {author} {\bibfnamefont {M.~J.}\ \bibnamefont {Sellars}},\ }\bibfield  {title} {\bibinfo {title} {Optically addressable nuclear spins in a solid with a six-hour coherence time},\ }\href {https://doi.org/10.1038/nature14025} {\bibfield  {journal} {\bibinfo  {journal} {Nature}\ }\textbf {\bibinfo {volume} {517}},\ \bibinfo {pages} {177} (\bibinfo {year} {2015})}\BibitemShut {NoStop}%
\bibitem [{\citenamefont {Rakonjac}\ \emph {et~al.}(2020)\citenamefont {Rakonjac}, \citenamefont {Chen}, \citenamefont {Horvath},\ and\ \citenamefont {Longdell}}]{Rakonjac_2020}%
  \BibitemOpen
  \bibfield  {author} {\bibinfo {author} {\bibfnamefont {J.~V.}\ \bibnamefont {Rakonjac}}, \bibinfo {author} {\bibfnamefont {Y.-H.}\ \bibnamefont {Chen}}, \bibinfo {author} {\bibfnamefont {S.~P.}\ \bibnamefont {Horvath}},\ and\ \bibinfo {author} {\bibfnamefont {J.~J.}\ \bibnamefont {Longdell}},\ }\bibfield  {title} {\bibinfo {title} {Long spin coherence times in the ground state and in an optically excited state of \textsuperscript{167}{Er}\textsuperscript{3+}{:Y}\textsubscript{2}{SiO}\textsubscript{5}},\ }\href {https://doi.org/10.1103/PhysRevB.101.184430} {\bibfield  {journal} {\bibinfo  {journal} {Physical Review B}\ }\textbf {\bibinfo {volume} {101}},\ \bibinfo {pages} {184430} (\bibinfo {year} {2020})}\BibitemShut {NoStop}%
\bibitem [{\citenamefont {Guo}\ \emph {et~al.}(2023)\citenamefont {Guo}, \citenamefont {Liu}, \citenamefont {Sun}, \citenamefont {Ren}, \citenamefont {Wang},\ and\ \citenamefont {Zhong}}]{Guo_2023}%
  \BibitemOpen
  \bibfield  {author} {\bibinfo {author} {\bibfnamefont {M.}~\bibnamefont {Guo}}, \bibinfo {author} {\bibfnamefont {S.}~\bibnamefont {Liu}}, \bibinfo {author} {\bibfnamefont {W.}~\bibnamefont {Sun}}, \bibinfo {author} {\bibfnamefont {M.}~\bibnamefont {Ren}}, \bibinfo {author} {\bibfnamefont {F.}~\bibnamefont {Wang}},\ and\ \bibinfo {author} {\bibfnamefont {M.}~\bibnamefont {Zhong}},\ }\bibfield  {title} {\bibinfo {title} {Rare-earth quantum memories: The experimental status quo},\ }\href {https://doi.org/10.1007/s11467-022-1240-8} {\bibfield  {journal} {\bibinfo  {journal} {Frontiers of Physics}\ }\textbf {\bibinfo {volume} {18}},\ \bibinfo {pages} {21303} (\bibinfo {year} {2023})}\BibitemShut {NoStop}%
\bibitem [{\citenamefont {Azuma}\ \emph {et~al.}(2023)\citenamefont {Azuma}, \citenamefont {Economou}, \citenamefont {Elkouss}, \citenamefont {Hilaire}, \citenamefont {Jiang}, \citenamefont {Lo},\ and\ \citenamefont {Tzitrin}}]{azuma_quantum_2023}%
  \BibitemOpen
  \bibfield  {author} {\bibinfo {author} {\bibfnamefont {K.}~\bibnamefont {Azuma}}, \bibinfo {author} {\bibfnamefont {S.~E.}\ \bibnamefont {Economou}}, \bibinfo {author} {\bibfnamefont {D.}~\bibnamefont {Elkouss}}, \bibinfo {author} {\bibfnamefont {P.}~\bibnamefont {Hilaire}}, \bibinfo {author} {\bibfnamefont {L.}~\bibnamefont {Jiang}}, \bibinfo {author} {\bibfnamefont {H.-K.}\ \bibnamefont {Lo}},\ and\ \bibinfo {author} {\bibfnamefont {I.}~\bibnamefont {Tzitrin}},\ }\bibfield  {title} {\bibinfo {title} {Quantum repeaters: {From} quantum networks to the quantum internet},\ }\href {https://doi.org/10.1103/RevModPhys.95.045006} {\bibfield  {journal} {\bibinfo  {journal} {Reviews of Modern Physics}\ }\textbf {\bibinfo {volume} {95}},\ \bibinfo {pages} {045006} (\bibinfo {year} {2023})}\BibitemShut {NoStop}%
\bibitem [{\citenamefont {Williamson}\ \emph {et~al.}(2014)\citenamefont {Williamson}, \citenamefont {Chen},\ and\ \citenamefont {Longdell}}]{Williamson_2014}%
  \BibitemOpen
  \bibfield  {author} {\bibinfo {author} {\bibfnamefont {L.~A.}\ \bibnamefont {Williamson}}, \bibinfo {author} {\bibfnamefont {Y.-H.}\ \bibnamefont {Chen}},\ and\ \bibinfo {author} {\bibfnamefont {J.~J.}\ \bibnamefont {Longdell}},\ }\bibfield  {title} {\bibinfo {title} {Magneto-optic modulator with unit quantum efficiency},\ }\href {https://doi.org/10.1103/PhysRevLett.113.203601} {\bibfield  {journal} {\bibinfo  {journal} {Physical Review Letters}\ }\textbf {\bibinfo {volume} {113}},\ \bibinfo {pages} {203601} (\bibinfo {year} {2014})}\BibitemShut {NoStop}%
\bibitem [{\citenamefont {Fernandez-Gonzalvo}\ \emph {et~al.}(2015)\citenamefont {Fernandez-Gonzalvo}, \citenamefont {Chen}, \citenamefont {Yin}, \citenamefont {Rogge},\ and\ \citenamefont {Longdell}}]{Fernandez_2015}%
  \BibitemOpen
  \bibfield  {author} {\bibinfo {author} {\bibfnamefont {X.}~\bibnamefont {Fernandez-Gonzalvo}}, \bibinfo {author} {\bibfnamefont {Y.-H.}\ \bibnamefont {Chen}}, \bibinfo {author} {\bibfnamefont {C.}~\bibnamefont {Yin}}, \bibinfo {author} {\bibfnamefont {S.}~\bibnamefont {Rogge}},\ and\ \bibinfo {author} {\bibfnamefont {J.~J.}\ \bibnamefont {Longdell}},\ }\bibfield  {title} {\bibinfo {title} {Coherent frequency up-conversion of microwaves to the optical telecommunications band in an {Er:YSO} crystal},\ }\href {https://doi.org/10.1103/PhysRevA.92.062313} {\bibfield  {journal} {\bibinfo  {journal} {Physical Review A}\ }\textbf {\bibinfo {volume} {92}},\ \bibinfo {pages} {062313} (\bibinfo {year} {2015})}\BibitemShut {NoStop}%
\bibitem [{\citenamefont {Everts}\ \emph {et~al.}(2019)\citenamefont {Everts}, \citenamefont {Berrington}, \citenamefont {Ahlefeldt},\ and\ \citenamefont {Longdell}}]{Everts_2019}%
  \BibitemOpen
  \bibfield  {author} {\bibinfo {author} {\bibfnamefont {J.~R.}\ \bibnamefont {Everts}}, \bibinfo {author} {\bibfnamefont {M.~C.}\ \bibnamefont {Berrington}}, \bibinfo {author} {\bibfnamefont {R.~L.}\ \bibnamefont {Ahlefeldt}},\ and\ \bibinfo {author} {\bibfnamefont {J.~J.}\ \bibnamefont {Longdell}},\ }\bibfield  {title} {\bibinfo {title} {Microwave to optical photon conversion via fully concentrated rare-earth-ion crystals},\ }\href {https://doi.org/10.1103/PhysRevA.99.063830} {\bibfield  {journal} {\bibinfo  {journal} {Physical Review A}\ }\textbf {\bibinfo {volume} {99}},\ \bibinfo {pages} {063830} (\bibinfo {year} {2019})}\BibitemShut {NoStop}%
\bibitem [{\citenamefont {Fernandez-Gonzalvo}\ \emph {et~al.}(2019)\citenamefont {Fernandez-Gonzalvo}, \citenamefont {Horvath}, \citenamefont {Chen},\ and\ \citenamefont {Longdell}}]{Fernandez-Gonzalvo_2019}%
  \BibitemOpen
  \bibfield  {author} {\bibinfo {author} {\bibfnamefont {X.}~\bibnamefont {Fernandez-Gonzalvo}}, \bibinfo {author} {\bibfnamefont {S.~P.}\ \bibnamefont {Horvath}}, \bibinfo {author} {\bibfnamefont {Y.-H.}\ \bibnamefont {Chen}},\ and\ \bibinfo {author} {\bibfnamefont {J.~J.}\ \bibnamefont {Longdell}},\ }\bibfield  {title} {\bibinfo {title} {Cavity-enhanced raman heterodyne spectroscopy in {Er$^{3+}$:Y$_2$SiO$_5$} for microwave to optical signal conversion},\ }\href {https://doi.org/10.1103/PhysRevA.100.033807} {\bibfield  {journal} {\bibinfo  {journal} {Phys. Rev. A}\ }\textbf {\bibinfo {volume} {100}},\ \bibinfo {pages} {033807} (\bibinfo {year} {2019})}\BibitemShut {NoStop}%
\bibitem [{\citenamefont {Barnett}\ and\ \citenamefont {Longdell}(2020)}]{Barnett_2020}%
  \BibitemOpen
  \bibfield  {author} {\bibinfo {author} {\bibfnamefont {P.~S.}\ \bibnamefont {Barnett}}\ and\ \bibinfo {author} {\bibfnamefont {J.~J.}\ \bibnamefont {Longdell}},\ }\bibfield  {title} {\bibinfo {title} {Theory of microwave-optical conversion using rare-earth-ion dopants},\ }\href {https://doi.org/10.1103/PhysRevA.102.063718} {\bibfield  {journal} {\bibinfo  {journal} {Physical Review A}\ }\textbf {\bibinfo {volume} {102}},\ \bibinfo {pages} {063718} (\bibinfo {year} {2020})}\BibitemShut {NoStop}%
\bibitem [{\citenamefont {Bartholomew}\ \emph {et~al.}(2020)\citenamefont {Bartholomew}, \citenamefont {Rochman}, \citenamefont {Xie}, \citenamefont {Kindem}, \citenamefont {Ruskuc}, \citenamefont {Craiciu}, \citenamefont {Lei},\ and\ \citenamefont {Faraon}}]{Bartholomew_2020}%
  \BibitemOpen
  \bibfield  {author} {\bibinfo {author} {\bibfnamefont {J.~G.}\ \bibnamefont {Bartholomew}}, \bibinfo {author} {\bibfnamefont {J.}~\bibnamefont {Rochman}}, \bibinfo {author} {\bibfnamefont {T.}~\bibnamefont {Xie}}, \bibinfo {author} {\bibfnamefont {J.~M.}\ \bibnamefont {Kindem}}, \bibinfo {author} {\bibfnamefont {A.}~\bibnamefont {Ruskuc}}, \bibinfo {author} {\bibfnamefont {I.}~\bibnamefont {Craiciu}}, \bibinfo {author} {\bibfnamefont {M.}~\bibnamefont {Lei}},\ and\ \bibinfo {author} {\bibfnamefont {A.}~\bibnamefont {Faraon}},\ }\bibfield  {title} {\bibinfo {title} {On-chip coherent microwave-to-optical transduction mediated by ytterbium in {YVO$_4$}},\ }\href {https://doi.org/10.1038/s41467-020-16996-x} {\bibfield  {journal} {\bibinfo  {journal} {Nature communications}\ }\textbf {\bibinfo {volume} {11}},\ \bibinfo {pages} {3266} (\bibinfo {year} {2020})}\BibitemShut {NoStop}%
\bibitem [{\citenamefont {King}\ \emph {et~al.}(2021)\citenamefont {King}, \citenamefont {Barnett}, \citenamefont {Bartholomew}, \citenamefont {Faraon},\ and\ \citenamefont {Longdell}}]{King_2021}%
  \BibitemOpen
  \bibfield  {author} {\bibinfo {author} {\bibfnamefont {G.~G.~G.}\ \bibnamefont {King}}, \bibinfo {author} {\bibfnamefont {P.~S.}\ \bibnamefont {Barnett}}, \bibinfo {author} {\bibfnamefont {J.~G.}\ \bibnamefont {Bartholomew}}, \bibinfo {author} {\bibfnamefont {A.}~\bibnamefont {Faraon}},\ and\ \bibinfo {author} {\bibfnamefont {J.~J.}\ \bibnamefont {Longdell}},\ }\bibfield  {title} {\bibinfo {title} {Probing strong coupling between a microwave cavity and a spin ensemble with raman heterodyne spectroscopy},\ }\href {https://doi.org/10.1103/PhysRevB.103.214305} {\bibfield  {journal} {\bibinfo  {journal} {Physical Review B}\ }\textbf {\bibinfo {volume} {103}},\ \bibinfo {pages} {214305} (\bibinfo {year} {2021})}\BibitemShut {NoStop}%
\bibitem [{\citenamefont {Xie}\ \emph {et~al.}(2021)\citenamefont {Xie}, \citenamefont {Rochman}, \citenamefont {Bartholomew}, \citenamefont {Ruskuc}, \citenamefont {Kindem}, \citenamefont {Craiciu}, \citenamefont {Thiel}, \citenamefont {Cone},\ and\ \citenamefont {Faraon}}]{Xie_2021}%
  \BibitemOpen
  \bibfield  {author} {\bibinfo {author} {\bibfnamefont {T.}~\bibnamefont {Xie}}, \bibinfo {author} {\bibfnamefont {J.}~\bibnamefont {Rochman}}, \bibinfo {author} {\bibfnamefont {J.~G.}\ \bibnamefont {Bartholomew}}, \bibinfo {author} {\bibfnamefont {A.}~\bibnamefont {Ruskuc}}, \bibinfo {author} {\bibfnamefont {J.~M.}\ \bibnamefont {Kindem}}, \bibinfo {author} {\bibfnamefont {I.}~\bibnamefont {Craiciu}}, \bibinfo {author} {\bibfnamefont {C.~W.}\ \bibnamefont {Thiel}}, \bibinfo {author} {\bibfnamefont {R.~L.}\ \bibnamefont {Cone}},\ and\ \bibinfo {author} {\bibfnamefont {A.}~\bibnamefont {Faraon}},\ }\bibfield  {title} {\bibinfo {title} {Characterisation of {Er$^{3+}$:YVO$_4$} for microwave to optical transduction},\ }\href {https://doi.org/10.1103/PhysRevB.104.054111} {\bibfield  {journal} {\bibinfo  {journal} {Phys. Rev. B}\ }\textbf {\bibinfo {volume} {104}},\ \bibinfo {pages} {054111} (\bibinfo {year} {2021})}\BibitemShut {NoStop}%
\bibitem [{\citenamefont {Lim}\ \emph {et~al.}(2024)\citenamefont {Lim}, \citenamefont {Choi},\ and\ \citenamefont {Hong}}]{Lim_2024}%
  \BibitemOpen
  \bibfield  {author} {\bibinfo {author} {\bibfnamefont {H.-J.}\ \bibnamefont {Lim}}, \bibinfo {author} {\bibfnamefont {G.}~\bibnamefont {Choi}},\ and\ \bibinfo {author} {\bibfnamefont {K.}~\bibnamefont {Hong}},\ }\bibfield  {title} {\bibinfo {title} {Optical and raman spectroscopies of \textsuperscript{171}{Yb}\textsuperscript{3+}:{Y}\textsubscript{2}{SiO}\textsubscript{5} hyperfine structure for application toward microwave-to-optical transducer},\ }\href {https://doi.org/10.1007/s40042-023-00975-8} {\bibfield  {journal} {\bibinfo  {journal} {Journal of the Korean Physical Society}\ }\textbf {\bibinfo {volume} {84}},\ \bibinfo {pages} {50} (\bibinfo {year} {2024})}\BibitemShut {NoStop}%
\bibitem [{\citenamefont {B\"ottger}\ \emph {et~al.}(2006)\citenamefont {B\"ottger}, \citenamefont {Thiel}, \citenamefont {Sun},\ and\ \citenamefont {Cone}}]{Bottger_2006_optical}%
  \BibitemOpen
  \bibfield  {author} {\bibinfo {author} {\bibfnamefont {T.}~\bibnamefont {B\"ottger}}, \bibinfo {author} {\bibfnamefont {C.~W.}\ \bibnamefont {Thiel}}, \bibinfo {author} {\bibfnamefont {Y.}~\bibnamefont {Sun}},\ and\ \bibinfo {author} {\bibfnamefont {R.~L.}\ \bibnamefont {Cone}},\ }\bibfield  {title} {\bibinfo {title} {Optical decoherence and spectral diffusion at $1.5\phantom{\rule{0.3em}{0ex}}\ensuremath{\mu}\mathrm{m}$ in {Er}\textsuperscript{3+}:{Y}\textsubscript{2}{SiO}\textsubscript{5} versus magnetic field, temperature, and {Er}\textsuperscript{3+} concentration},\ }\href {https://doi.org/10.1103/PhysRevB.73.075101} {\bibfield  {journal} {\bibinfo  {journal} {Physical Review B}\ }\textbf {\bibinfo {volume} {73}},\ \bibinfo {pages} {075101} (\bibinfo {year} {2006})}\BibitemShut {NoStop}%
\bibitem [{\citenamefont {Ahlefeldt}\ \emph {et~al.}(2016)\citenamefont {Ahlefeldt}, \citenamefont {Hush},\ and\ \citenamefont {Sellars}}]{Ahlefeldt_2016}%
  \BibitemOpen
  \bibfield  {author} {\bibinfo {author} {\bibfnamefont {R.}~\bibnamefont {Ahlefeldt}}, \bibinfo {author} {\bibfnamefont {M.~R.}\ \bibnamefont {Hush}},\ and\ \bibinfo {author} {\bibfnamefont {M.}~\bibnamefont {Sellars}},\ }\bibfield  {title} {\bibinfo {title} {Ultranarrow optical inhomogeneous linewidth in a stoichiometric rare-earth crystal},\ }\href {https://doi.org/10.1103/PhysRevLett.117.250504} {\bibfield  {journal} {\bibinfo  {journal} {Physical review letters}\ }\textbf {\bibinfo {volume} {117}},\ \bibinfo {pages} {250504} (\bibinfo {year} {2016})}\BibitemShut {NoStop}%
\bibitem [{\citenamefont {Ahlefeldt}\ \emph {et~al.}(2013{\natexlab{a}})\citenamefont {Ahlefeldt}, \citenamefont {Manson},\ and\ \citenamefont {Sellars}}]{ahlefeldt_optical_2013}%
  \BibitemOpen
  \bibfield  {author} {\bibinfo {author} {\bibfnamefont {R.~L.}\ \bibnamefont {Ahlefeldt}}, \bibinfo {author} {\bibfnamefont {N.~B.}\ \bibnamefont {Manson}},\ and\ \bibinfo {author} {\bibfnamefont {M.~J.}\ \bibnamefont {Sellars}},\ }\bibfield  {title} {\bibinfo {title} {Optical lifetime and linewidth studies of the \textsuperscript{7}{F}\textsubscript{0} $\rightarrow$ \textsuperscript{5}{D}\textsubscript{0} transition in {EuCl}\textsubscript{3} $\cdot$ {6H}\textsubscript{2}{O}: {A} potential material for quantum memory applications},\ }\href {https://doi.org/10.1016/j.jlumin.2011.12.036} {\bibfield  {journal} {\bibinfo  {journal} {Journal of Luminescence}\ }\textbf {\bibinfo {volume} {133}},\ \bibinfo {pages} {152} (\bibinfo {year} {2013}{\natexlab{a}})}\BibitemShut {NoStop}%
\bibitem [{\citenamefont {Berrington}(2022)}]{Berrington_2022}%
  \BibitemOpen
  \bibfield  {author} {\bibinfo {author} {\bibfnamefont {M.~C.}\ \bibnamefont {Berrington}},\ }\emph {\bibinfo {title} {Optical studies of magnetically ordered erbium crystals}},\ \href {https://doi.org/10.25911/ZC9F-WA62} {\bibinfo {type} {{Ph.D.} thesis}},\ \bibinfo  {school} {Australian National University} (\bibinfo {year} {2022})\BibitemShut {NoStop}%
\bibitem [{\citenamefont {Kukharchyk}\ \emph {et~al.}(2018)\citenamefont {Kukharchyk}, \citenamefont {Sholokhov}, \citenamefont {Morozov}, \citenamefont {Korableva}, \citenamefont {Kalachev},\ and\ \citenamefont {Bushev}}]{kukharchyk_optical_2018}%
  \BibitemOpen
  \bibfield  {author} {\bibinfo {author} {\bibfnamefont {N.}~\bibnamefont {Kukharchyk}}, \bibinfo {author} {\bibfnamefont {D.}~\bibnamefont {Sholokhov}}, \bibinfo {author} {\bibfnamefont {O.}~\bibnamefont {Morozov}}, \bibinfo {author} {\bibfnamefont {S.~L.}\ \bibnamefont {Korableva}}, \bibinfo {author} {\bibfnamefont {A.~A.}\ \bibnamefont {Kalachev}},\ and\ \bibinfo {author} {\bibfnamefont {P.~A.}\ \bibnamefont {Bushev}},\ }\bibfield  {title} {\bibinfo {title} {Optical coherence of \textsuperscript{166}{Er}:\textsuperscript{7}{LiYF}\textsubscript{4} crystal below 1 {K}},\ }\href {https://doi.org/10.1088/1367-2630/aaa7e4} {\bibfield  {journal} {\bibinfo  {journal} {New Journal of Physics}\ }\textbf {\bibinfo {volume} {20}},\ \bibinfo {pages} {023044} (\bibinfo {year} {2018})}\BibitemShut {NoStop}%
\bibitem [{\citenamefont {Hiraishi}\ \emph {et~al.}(2025)\citenamefont {Hiraishi}, \citenamefont {Roberts}, \citenamefont {King}, \citenamefont {Trainor},\ and\ \citenamefont {Longdell}}]{hiraishi_long_2025}%
  \BibitemOpen
  \bibfield  {author} {\bibinfo {author} {\bibfnamefont {M.}~\bibnamefont {Hiraishi}}, \bibinfo {author} {\bibfnamefont {Z.~H.}\ \bibnamefont {Roberts}}, \bibinfo {author} {\bibfnamefont {G.~G.~G.}\ \bibnamefont {King}}, \bibinfo {author} {\bibfnamefont {L.~S.}\ \bibnamefont {Trainor}},\ and\ \bibinfo {author} {\bibfnamefont {J.~J.}\ \bibnamefont {Longdell}},\ }\bibfield  {title} {\bibinfo {title} {Long optical coherence times in a rare-earth-doped antiferromagnet},\ }\href {https://doi.org/10.1038/s41567-025-02920-x} {\bibfield  {journal} {\bibinfo  {journal} {Nature Physics}\ }\textbf {\bibinfo {volume} {21}},\ \bibinfo {pages} {1112} (\bibinfo {year} {2025})}\BibitemShut {NoStop}%
\bibitem [{\citenamefont {Li}\ \emph {et~al.}(2020)\citenamefont {Li}, \citenamefont {Huang}, \citenamefont {Zhu}, \citenamefont {Liu}, \citenamefont {Liu}, \citenamefont {Zhou}, \citenamefont {Li},\ and\ \citenamefont {Guo}}]{li_optical_2020}%
  \BibitemOpen
  \bibfield  {author} {\bibinfo {author} {\bibfnamefont {P.-Y.}\ \bibnamefont {Li}}, \bibinfo {author} {\bibfnamefont {J.-Y.}\ \bibnamefont {Huang}}, \bibinfo {author} {\bibfnamefont {T.-X.}\ \bibnamefont {Zhu}}, \bibinfo {author} {\bibfnamefont {C.}~\bibnamefont {Liu}}, \bibinfo {author} {\bibfnamefont {D.-C.}\ \bibnamefont {Liu}}, \bibinfo {author} {\bibfnamefont {Z.-Q.}\ \bibnamefont {Zhou}}, \bibinfo {author} {\bibfnamefont {C.-F.}\ \bibnamefont {Li}},\ and\ \bibinfo {author} {\bibfnamefont {G.-C.}\ \bibnamefont {Guo}},\ }\bibfield  {title} {\bibinfo {title} {Optical spectroscopy and coherent dynamics of \textsuperscript{167}{Er}\textsuperscript{3+}:{YVO}\textsubscript{4} at 1.5 $\mu$m below 1 {K}},\ }\href {https://doi.org/10.1016/j.jlumin.2020.117344} {\bibfield  {journal} {\bibinfo  {journal} {Journal of Luminescence}\ }\textbf {\bibinfo {volume} {225}},\ \bibinfo {pages} {117344} (\bibinfo {year} {2020})}\BibitemShut {NoStop}%
\bibitem [{\citenamefont {Marti}\ \emph {et~al.}(1995)\citenamefont {Marti}, \citenamefont {Medarde}, \citenamefont {Rosenkranz}, \citenamefont {Fischer}, \citenamefont {Furrer},\ and\ \citenamefont {Klemenz}}]{Marti1995}%
  \BibitemOpen
  \bibfield  {author} {\bibinfo {author} {\bibfnamefont {W.}~\bibnamefont {Marti}}, \bibinfo {author} {\bibfnamefont {M.}~\bibnamefont {Medarde}}, \bibinfo {author} {\bibfnamefont {S.}~\bibnamefont {Rosenkranz}}, \bibinfo {author} {\bibfnamefont {P.}~\bibnamefont {Fischer}}, \bibinfo {author} {\bibfnamefont {A.}~\bibnamefont {Furrer}},\ and\ \bibinfo {author} {\bibfnamefont {C.}~\bibnamefont {Klemenz}},\ }\bibfield  {title} {\bibinfo {title} {Hyperfine-enhanced nuclear polarization in {NdGaO$_3$}},\ }\href {https://doi.org/10.1103/PhysRevB.52.4275} {\bibfield  {journal} {\bibinfo  {journal} {Phys. Rev. B}\ }\textbf {\bibinfo {volume} {52}},\ \bibinfo {pages} {4275} (\bibinfo {year} {1995})}\BibitemShut {NoStop}%
\bibitem [{\citenamefont {Luis}\ \emph {et~al.}(1998)\citenamefont {Luis}, \citenamefont {Kuz'min}, \citenamefont {Bartolom\'e}, \citenamefont {Orera}, \citenamefont {Bartolom\'e}, \citenamefont {Artigas},\ and\ \citenamefont {Rub\'{\i}n}}]{Luis1998}%
  \BibitemOpen
  \bibfield  {author} {\bibinfo {author} {\bibfnamefont {F.}~\bibnamefont {Luis}}, \bibinfo {author} {\bibfnamefont {M.~D.}\ \bibnamefont {Kuz'min}}, \bibinfo {author} {\bibfnamefont {F.}~\bibnamefont {Bartolom\'e}}, \bibinfo {author} {\bibfnamefont {V.~M.}\ \bibnamefont {Orera}}, \bibinfo {author} {\bibfnamefont {J.}~\bibnamefont {Bartolom\'e}}, \bibinfo {author} {\bibfnamefont {M.}~\bibnamefont {Artigas}},\ and\ \bibinfo {author} {\bibfnamefont {J.}~\bibnamefont {Rub\'{\i}n}},\ }\bibfield  {title} {\bibinfo {title} {Magnetic susceptibility of {NdGaO$_3$} at low temperatures: A quasi-two-dimensional {Ising} behavior},\ }\href {https://doi.org/10.1103/PhysRevB.58.798} {\bibfield  {journal} {\bibinfo  {journal} {Phys. Rev. B}\ }\textbf {\bibinfo {volume} {58}},\ \bibinfo {pages} {798} (\bibinfo {year} {1998})}\BibitemShut {NoStop}%
\bibitem [{\citenamefont {Koster}(1963)}]{Koster_1963}%
  \BibitemOpen
  \bibfield  {author} {\bibinfo {author} {\bibfnamefont {G.~F.}\ \bibnamefont {Koster}},\ }\bibfield  {title} {\bibinfo {title} {Properties of the thirty-two point groups},\ }\href@noop {} {\bibfield  {journal} {\bibinfo  {journal} {MIT Press, Cambridge, MA, USA}\ } (\bibinfo {year} {1963})}\BibitemShut {NoStop}%
\bibitem [{\citenamefont {Orera}\ \emph {et~al.}(1995{\natexlab{a}})\citenamefont {Orera}, \citenamefont {Trinkler}, \citenamefont {Merino},\ and\ \citenamefont {Larrea}}]{Orera1995_1}%
  \BibitemOpen
  \bibfield  {author} {\bibinfo {author} {\bibfnamefont {V.~M.}\ \bibnamefont {Orera}}, \bibinfo {author} {\bibfnamefont {L.~E.}\ \bibnamefont {Trinkler}}, \bibinfo {author} {\bibfnamefont {R.~I.}\ \bibnamefont {Merino}},\ and\ \bibinfo {author} {\bibfnamefont {A.}~\bibnamefont {Larrea}},\ }\bibfield  {title} {\bibinfo {title} {The optical properties of the {Nd$^{3+}$} ion in {NdGaO$_3$} and {LaGaO$_3$:Nd}: temperature and concentration dependence},\ }\href {https://doi.org/10.1088/0953-8984/7/49/027} {\bibfield  {journal} {\bibinfo  {journal} {Journal of Physics: Condensed Matter}\ }\textbf {\bibinfo {volume} {7}},\ \bibinfo {pages} {9657} (\bibinfo {year} {1995}{\natexlab{a}})}\BibitemShut {NoStop}%
\bibitem [{\citenamefont {Orera}\ \emph {et~al.}(1995{\natexlab{b}})\citenamefont {Orera}, \citenamefont {Trinkler},\ and\ \citenamefont {Merino}}]{Orera1995_2}%
  \BibitemOpen
  \bibfield  {author} {\bibinfo {author} {\bibfnamefont {V.~M.}\ \bibnamefont {Orera}}, \bibinfo {author} {\bibfnamefont {L.~E.}\ \bibnamefont {Trinkler}},\ and\ \bibinfo {author} {\bibfnamefont {R.~I.}\ \bibnamefont {Merino}},\ }\bibfield  {title} {\bibinfo {title} {Differential spectroscopic properties of {Nd$^{3+}$} in {NdGaO$_3$} and {LaGaO$_3$}},\ }\href {https://doi.org/10.1080/10420159508229829} {\bibfield  {journal} {\bibinfo  {journal} {Radiation Effects and Defects in Solids}\ }\textbf {\bibinfo {volume} {135}},\ \bibinfo {pages} {173} (\bibinfo {year} {1995}{\natexlab{b}})}\BibitemShut {NoStop}%
\bibitem [{\citenamefont {Carnall}\ \emph {et~al.}(1989)\citenamefont {Carnall}, \citenamefont {Goodman}, \citenamefont {Rajnak},\ and\ \citenamefont {Rana}}]{Carnall_1989}%
  \BibitemOpen
  \bibfield  {author} {\bibinfo {author} {\bibfnamefont {W.~T.}\ \bibnamefont {Carnall}}, \bibinfo {author} {\bibfnamefont {G.~L.}\ \bibnamefont {Goodman}}, \bibinfo {author} {\bibfnamefont {K.}~\bibnamefont {Rajnak}},\ and\ \bibinfo {author} {\bibfnamefont {R.~S.}\ \bibnamefont {Rana}},\ }\bibfield  {title} {\bibinfo {title} {A systematic analysis of the spectra of the lanthanides doped into single crystal {LaF}\textsubscript{3}},\ }\href {https://doi.org/10.1063/1.455853} {\bibfield  {journal} {\bibinfo  {journal} {The Journal of Chemical Physics}\ }\textbf {\bibinfo {volume} {90}},\ \bibinfo {pages} {3443} (\bibinfo {year} {1989})}\BibitemShut {NoStop}%
\bibitem [{\citenamefont {Nov\'ak}\ \emph {et~al.}(2013)\citenamefont {Nov\'ak}, \citenamefont {Kn{\'{\i}}\v{z}ek}, \citenamefont {Mary\v{s}ko}, \citenamefont {Jir\'{a}k},\ and\ \citenamefont {Kune\v{s}}}]{Novak2013}%
  \BibitemOpen
  \bibfield  {author} {\bibinfo {author} {\bibfnamefont {P.}~\bibnamefont {Nov\'ak}}, \bibinfo {author} {\bibfnamefont {K.}~\bibnamefont {Kn{\'{\i}}\v{z}ek}}, \bibinfo {author} {\bibfnamefont {M.}~\bibnamefont {Mary\v{s}ko}}, \bibinfo {author} {\bibfnamefont {Z.}~\bibnamefont {Jir\'{a}k}},\ and\ \bibinfo {author} {\bibfnamefont {J.}~\bibnamefont {Kune\v{s}}},\ }\bibfield  {title} {\bibinfo {title} {Crystal field and magnetism of {Pr$^{3+}$} and {Nd$^{3+}$} ions in orthorhombic perovskites},\ }\href {https://doi.org/10.1088/0953-8984/25/44/446001} {\bibfield  {journal} {\bibinfo  {journal} {Journal of Physics: Condensed Matter}\ }\textbf {\bibinfo {volume} {25}},\ \bibinfo {pages} {446001} (\bibinfo {year} {2013})}\BibitemShut {NoStop}%
\bibitem [{die()}]{dieke}%
  \BibitemOpen
  \href@noop {} {}\bibinfo {howpublished} {\href{https://github.com/jevonlongdell/dieke}{https://github.com/jevonlongdell/dieke}}\BibitemShut {NoStop}%
\bibitem [{\citenamefont {Guillot-No\"el}\ \emph {et~al.}(2005)\citenamefont {Guillot-No\"el}, \citenamefont {Goldner}, \citenamefont {Antic-Fidancev},\ and\ \citenamefont {Le~Gou\"et}}]{Guillot_2005}%
  \BibitemOpen
  \bibfield  {author} {\bibinfo {author} {\bibfnamefont {O.}~\bibnamefont {Guillot-No\"el}}, \bibinfo {author} {\bibfnamefont {P.}~\bibnamefont {Goldner}}, \bibinfo {author} {\bibfnamefont {E.}~\bibnamefont {Antic-Fidancev}},\ and\ \bibinfo {author} {\bibfnamefont {J.~L.}\ \bibnamefont {Le~Gou\"et}},\ }\bibfield  {title} {\bibinfo {title} {Analysis of magnetic interactions in rare-earth-doped crystals for quantum manipulation},\ }\href {https://doi.org/10.1103/PhysRevB.71.174409} {\bibfield  {journal} {\bibinfo  {journal} {Physical Review B}\ }\textbf {\bibinfo {volume} {71}},\ \bibinfo {pages} {174409} (\bibinfo {year} {2005})}\BibitemShut {NoStop}%
\bibitem [{\citenamefont {Afzelius}\ \emph {et~al.}(2010)\citenamefont {Afzelius}, \citenamefont {Staudt}, \citenamefont {{de Riedmatten}}, \citenamefont {Gisin}, \citenamefont {Guillot-No\"el}, \citenamefont {Goldner}, \citenamefont {Marino}, \citenamefont {Porcher}, \citenamefont {Cavalli},\ and\ \citenamefont {Bettinelli}}]{Afzelius_2010}%
  \BibitemOpen
  \bibfield  {author} {\bibinfo {author} {\bibfnamefont {M.}~\bibnamefont {Afzelius}}, \bibinfo {author} {\bibfnamefont {M.~U.}\ \bibnamefont {Staudt}}, \bibinfo {author} {\bibfnamefont {H.}~\bibnamefont {{de Riedmatten}}}, \bibinfo {author} {\bibfnamefont {N.}~\bibnamefont {Gisin}}, \bibinfo {author} {\bibfnamefont {O.}~\bibnamefont {Guillot-No\"el}}, \bibinfo {author} {\bibfnamefont {P.}~\bibnamefont {Goldner}}, \bibinfo {author} {\bibfnamefont {R.}~\bibnamefont {Marino}}, \bibinfo {author} {\bibfnamefont {P.}~\bibnamefont {Porcher}}, \bibinfo {author} {\bibfnamefont {E.}~\bibnamefont {Cavalli}},\ and\ \bibinfo {author} {\bibfnamefont {M.}~\bibnamefont {Bettinelli}},\ }\bibfield  {title} {\bibinfo {title} {Efficient optical pumping of {Zeeman} spin levels in {Nd$^{3+}$:YVO$_4$}},\ }\href {https://doi.org/https://doi.org/10.1016/j.jlumin.2009.12.026} {\bibfield  {journal} {\bibinfo  {journal} {Journal of Luminescence}\ }\textbf {\bibinfo {volume} {130}},\ \bibinfo {pages} {1566} (\bibinfo {year}
  {2010})}\BibitemShut {NoStop}%
\bibitem [{\citenamefont {Ekin}(2006)}]{Ekin_2006}%
  \BibitemOpen
  \bibfield  {author} {\bibinfo {author} {\bibfnamefont {J.}~\bibnamefont {Ekin}},\ }\href {https://doi.org/10.1093/acprof:oso/9780198570547.001.0001} {\emph {\bibinfo {title} {{Experimental Techniques for Low-Temperature Measurements: Cryostat Design, Material Properties and Superconductor Critical-Current Testing}}}}\ (\bibinfo  {publisher} {Oxford University Press},\ \bibinfo {year} {2006})\BibitemShut {NoStop}%
\bibitem [{\citenamefont {Cone}\ \emph {et~al.}(1993)\citenamefont {Cone}, \citenamefont {Hansen}, \citenamefont {Leask},\ and\ \citenamefont {Wanklyn}}]{Cone_1993}%
  \BibitemOpen
  \bibfield  {author} {\bibinfo {author} {\bibfnamefont {R.~L.}\ \bibnamefont {Cone}}, \bibinfo {author} {\bibfnamefont {P.~C.}\ \bibnamefont {Hansen}}, \bibinfo {author} {\bibfnamefont {M.~J.~M.}\ \bibnamefont {Leask}},\ and\ \bibinfo {author} {\bibfnamefont {B.}~\bibnamefont {Wanklyn}},\ }\bibfield  {title} {\bibinfo {title} {Optical fluorescence excitation spectra of flux-grown stoichiometric europium vanadate crystals},\ }\href {https://doi.org/10.1088/0953-8984/5/5/009} {\bibfield  {journal} {\bibinfo  {journal} {Journal of Physics: Condensed Matter}\ }\textbf {\bibinfo {volume} {5}},\ \bibinfo {pages} {573} (\bibinfo {year} {1993})}\BibitemShut {NoStop}%
\bibitem [{\citenamefont {Jaaniso}\ \emph {et~al.}(1994)\citenamefont {Jaaniso}, \citenamefont {Hagemann},\ and\ \citenamefont {Bill}}]{Jaaniso1994}%
  \BibitemOpen
  \bibfield  {author} {\bibinfo {author} {\bibfnamefont {R.}~\bibnamefont {Jaaniso}}, \bibinfo {author} {\bibfnamefont {H.}~\bibnamefont {Hagemann}},\ and\ \bibinfo {author} {\bibfnamefont {H.}~\bibnamefont {Bill}},\ }\bibfield  {title} {\bibinfo {title} {{Inhomogeneous broadening of optical spectra in mixed crystals: Basic model and its application to Sm$^{2+}$ in SrFCl$_{x}$Br$_{1-x}$}},\ }\href {https://doi.org/10.1063/1.467912} {\bibfield  {journal} {\bibinfo  {journal} {The Journal of Chemical Physics}\ }\textbf {\bibinfo {volume} {101}},\ \bibinfo {pages} {10323} (\bibinfo {year} {1994})}\BibitemShut {NoStop}%
\bibitem [{\citenamefont {Yamaguchi}\ \emph {et~al.}(1999)\citenamefont {Yamaguchi}, \citenamefont {Koyama}, \citenamefont {Suemoto},\ and\ \citenamefont {Mitsunaga}}]{Yamaguchi1999}%
  \BibitemOpen
  \bibfield  {author} {\bibinfo {author} {\bibfnamefont {M.}~\bibnamefont {Yamaguchi}}, \bibinfo {author} {\bibfnamefont {K.}~\bibnamefont {Koyama}}, \bibinfo {author} {\bibfnamefont {T.}~\bibnamefont {Suemoto}},\ and\ \bibinfo {author} {\bibfnamefont {M.}~\bibnamefont {Mitsunaga}},\ }\bibfield  {title} {\bibinfo {title} {Perturbed ion sites in {Eu$^{3+}$YAlO$_3$} studied by optical-rf double-resonance spectroscopy},\ }\href {https://doi.org/10.1103/PhysRevB.59.9126} {\bibfield  {journal} {\bibinfo  {journal} {Physical Review B}\ }\textbf {\bibinfo {volume} {59}},\ \bibinfo {pages} {9126} (\bibinfo {year} {1999})}\BibitemShut {NoStop}%
\bibitem [{\citenamefont {Ahlefeldt}\ \emph {et~al.}(2013{\natexlab{b}})\citenamefont {Ahlefeldt}, \citenamefont {Hutchison}, \citenamefont {Manson},\ and\ \citenamefont {Sellars}}]{Ahlefeldt2013}%
  \BibitemOpen
  \bibfield  {author} {\bibinfo {author} {\bibfnamefont {R.~L.}\ \bibnamefont {Ahlefeldt}}, \bibinfo {author} {\bibfnamefont {W.~D.}\ \bibnamefont {Hutchison}}, \bibinfo {author} {\bibfnamefont {N.~B.}\ \bibnamefont {Manson}},\ and\ \bibinfo {author} {\bibfnamefont {M.~J.}\ \bibnamefont {Sellars}},\ }\bibfield  {title} {\bibinfo {title} {Method for assigning satellite lines to crystallographic sites in rare-earth crystals},\ }\href {https://doi.org/10.1103/PhysRevB.88.184424} {\bibfield  {journal} {\bibinfo  {journal} {Physical Review B}\ }\textbf {\bibinfo {volume} {88}},\ \bibinfo {pages} {184424} (\bibinfo {year} {2013}{\natexlab{b}})}\BibitemShut {NoStop}%
\bibitem [{\citenamefont {Blaauw-Smith}\ \emph {et~al.}(2024)\citenamefont {Blaauw-Smith}, \citenamefont {Trainor}, \citenamefont {King}, \citenamefont {Lambert}, \citenamefont {Hiraishi},\ and\ \citenamefont {Longdell}}]{blaauw-smith_magnetostriction_2024}%
  \BibitemOpen
  \bibfield  {author} {\bibinfo {author} {\bibfnamefont {F.~J.}\ \bibnamefont {Blaauw-Smith}}, \bibinfo {author} {\bibfnamefont {L.~S.}\ \bibnamefont {Trainor}}, \bibinfo {author} {\bibfnamefont {G.~G.~G.}\ \bibnamefont {King}}, \bibinfo {author} {\bibfnamefont {N.~J.}\ \bibnamefont {Lambert}}, \bibinfo {author} {\bibfnamefont {M.}~\bibnamefont {Hiraishi}},\ and\ \bibinfo {author} {\bibfnamefont {J.~J.}\ \bibnamefont {Longdell}},\ }\bibfield  {title} {\bibinfo {title} {Magnetostriction measurements at milli-kelvin temperatures using a {Fabry}–{Pérot} interferometer},\ }\href {https://doi.org/10.1063/5.0191294} {\bibfield  {journal} {\bibinfo  {journal} {Review of Scientific Instruments}\ }\textbf {\bibinfo {volume} {95}},\ \bibinfo {pages} {043905} (\bibinfo {year} {2024})}\BibitemShut {NoStop}%
\bibitem [{\citenamefont {Danisch}\ and\ \citenamefont {Krumbiegel}(2021)}]{DanischKrumbiegel2021}%
  \BibitemOpen
  \bibfield  {author} {\bibinfo {author} {\bibfnamefont {S.}~\bibnamefont {Danisch}}\ and\ \bibinfo {author} {\bibfnamefont {J.}~\bibnamefont {Krumbiegel}},\ }\bibfield  {title} {\bibinfo {title} {{Makie.jl}: Flexible high-performance data visualization for {Julia}},\ }\href {https://doi.org/10.21105/joss.03349} {\bibfield  {journal} {\bibinfo  {journal} {Journal of Open Source Software}\ }\textbf {\bibinfo {volume} {6}},\ \bibinfo {pages} {3349} (\bibinfo {year} {2021})}\BibitemShut {NoStop}%
\end{thebibliography}

\end{document}